\documentclass[prd,aps,twocolumn,showpacs,superscriptaddress,floatfix]{revtex4-1}
\usepackage{setspace}
\usepackage{bm}
\usepackage{pdfpages}
\usepackage{amssymb}
\usepackage{booktabs}
\usepackage{amsmath}
\usepackage{txfonts}
\usepackage{empheq}
\usepackage{graphicx}% Include figure files
\usepackage{dcolumn}% Align table columns on decimal point
\usepackage{bm}% bold math
\usepackage{color}
\usepackage{tikz}
\usetikzlibrary{shapes,backgrounds} 
\def\comment#1{}

\usepackage{amsmath}
\def\slashchar#1{\setbox0=\hbox{$#1$}           % set a box for #1
   \dimen0=\wd0                                 % and get its size
   \setbox1=\hbox{/} \dimen1=\wd1               % get size of /
   \ifdim\dimen0>\dimen1                        % #1 is bigger
      \rlap{\hbox to \dimen0{\hfil/\hfil}}      % so center / in box
      #1                                        % and print #1
   \else                                        % / is bigger
      \rlap{\hbox to \dimen1{\hfil$#1$\hfil}}   % so center #1
      /                                         % and print /
   \fi}                                         %

\begin{document}

%\title{Relaxation time bounds and possible sharp quantum signatures in ``classical'' fluids}
%\title{Glassy dynamics and thermodynamics as a quantum corollary of the liquid to solid transition}
%\title{The ``glass transition'' as a quantum corollary of the liquid to solid transition}
%\title{Glassy dynamics as a quantum corollary of the liquid to solid transition and a single parameter fit}
\title{A one parameter fit for glassy dynamics as a quantum corollary of the liquid to solid transition}
%\title{A one parameter fit for glassy dynamics as a corollary of the equilibrium liquid to solid transition}

\author{Zohar Nussinov}
\affiliation{Department of Physics, Washington University, St. Louis,
MO 63130, U.S.A.}
\email{zohar@wuphys.wustl.edu}

\date{\today}

 \begin{abstract}
We apply microcanonical ensemble considerations to suggest that, whenever it may thermalize, a general disorder-free many-body Hamiltonian of a typical atomic system has solid-like eigenstates at low energies and fluid-type (and gaseous, plasma) eigenstates associated with energy densities exceeding those present in the melting (and, respectively, higher energy) transition(s). In particular, the lowest energy density at which the eigenstates of such a clean many body atomic system undergo a non-analytic change is that of the melting (or freezing) transition. We invoke this observation to analyze the evolution of a liquid upon supercooling (i.e., cooling rapidly enough to avoid solidification below the freezing temperature). Expanding the wavefunction of a supercooled liquid in the complete eigenbasis of the many-body Hamiltonian, only the higher energy liquid-type eigenstates contribute significantly to measurable hydrodynamic relaxations (e.g., those probed by viscosity)
while static thermodynamic observables become weighted averages over both solid- and liquid-type eigenstates. Consequently, when extrapolated to low temperatures, hydrodynamic relaxation times of deeply supercooled liquids (i.e., glasses) may seem to diverge at nearly the same temperature at 
which the extrapolated entropy of the supercooled liquid becomes that of the solid. In this formal quantum framework, the increasingly sluggish (and spatially heterogeneous) dynamics 
in supercooled liquids as their temperature is lowered stems from the existence of the single non-analytic change of the eigenstates of the clean many-body Hamiltonian
at the equilibrium melting transition present in low energy solid-type eigenstates. We derive a {\it single} (possibly computable) {\it  dimensionless parameter} fit to the viscosity and suggest other testable predictions of our approach. 
\end{abstract}


\maketitle

\section{Introduction}
\begin{figure*}
\centering
\includegraphics[width=2.15 \columnwidth]{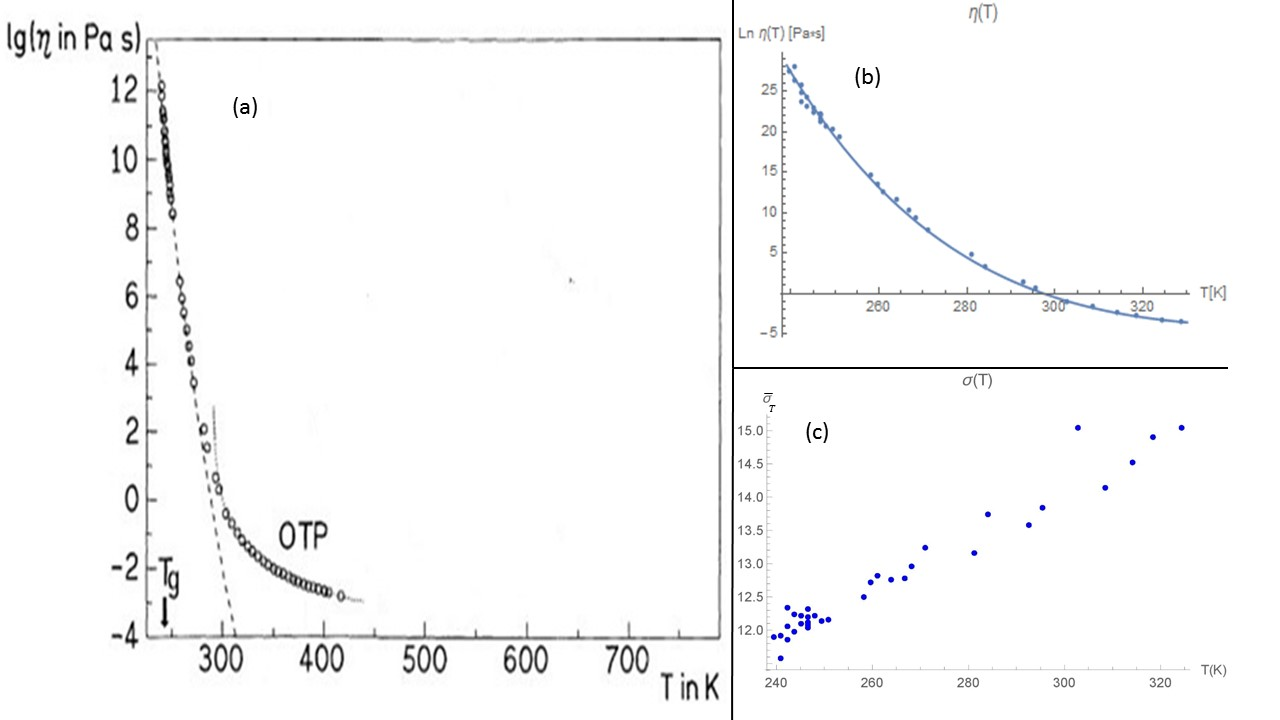}
\caption{(a) Published viscosity data (and earlier attempted fits)  \cite{rs} of o-terphenyl (OTP), a quintessential ``fragile'' glass former. (A ``fragile'' liquid \cite{fragile1} is one
in which the viscosity increases dramatically at low temperatures.) $T_g$ is the glass transition temperature (defined as the temperature at which $\eta(T_g) = 10^{12} ~$Pa $\cdot$ s).
(b) Result of a simplified rendition of our theory provided by Eq. (\ref{geoni}). This fit has only a {\it single dimensionless parameter}.
In Eq. (\ref{geoni}), we set the melting temperature to its experimental value $T_m$=329.35 K and
the effective width to be $\overline{\sigma}_T = \overline{A}T$ with the single fitting parameter $\overline{A} \approx  0.049$.
(c) A numerical evaluation of the effective width $\overline{\sigma}_T$ when equating Eq. (\ref{geoni}) to the 
experimental value in panel (a). The linear increase in $\overline{\sigma}_T$ in $T$ is manifest in (c). }
\label{OTP.}
\end{figure*}

The enigmatic ``glass transition'' \cite{paw,kt} appears in nearly all liquids. 
The basic observation (employed for millennia) is that liquids may be cooled sufficiently rapidly
(or, so-called, ``supercooled'') past their freezing temperatures so that they bypass crystallization and at sufficiently low temperatures ($T<T_{g}$) form an amorphous solid like material, a ``glass'' \cite{define-glass}. This ``transition'' into a glass differs significantly from conventional thermodynamic transitions in many ways. Perhaps most notable is the disparity between dynamic and thermodynamic features. It is not uncommon to find a huge, e.g., $10^{14}$-fold, increase in the relaxation time as the temperature of a supercooled liquid is dropped \cite{fragile1,fragile2} (see, e.g., panel (a) of Fig. \ref{OTP.}). However, such a spectacular change in the dynamics is not accompanied by matching sizable changes in thermodynamic measurements such as that of the specific heat. The standard classical descriptions of glasses are hampered by the existence of many competing low energy states. 

 There have been many penetrating works that provided illuminating ideas (see, e.g., \cite{ag,ag1,ag2,ag3,CG,ejcg,elm,jam,jam',gt1,sri,gt2,gt3,gt4,gt5,gt6,moore,myega,wg,mdd,avoided1,avoided2,avoided3} for only a small subset of a very vast array) to address this question. Collectively, the concepts that these works were of great utility in numerous fields. However, no model is currently universally accepted.
%This complexity was incorporated
%in myriad classical physics based theories.  %that was masterfully employed in some 
%of these and other theories.
The most celebrated fit for the viscosity of
supercooled liquids (claimed by most, yet not all \cite{CG,ejcg,elm,myega,mdd,avoided1,avoided2,avoided3} theories) is the Vogel-Fulcher-Tammann-Hesse (VFTH) fit \cite{vft1,vft2,vft3,vft4,vft5},
%\begin{eqnarray}
%\label{VFT}
$\eta_{VFTH} = \eta_{0} e^{DT_{0}/(T-T_{0})}$,
%\end{eqnarray}
where $\eta_{0}, D,$ and $T_{0}(<T_{g})$ are liquid dependent constants.
Thus, the VFTH function 
asserts that at $T=T_{0}$
the viscosity (and relaxation times) diverge. The VFTH and similar fits have come under experimental scrutiny, e.g., \cite{critical_vft}. Theories deriving the VFTH function and most others imply various special temperatures. Our approach to the problem is different. We suggest that genuine phase transitions must coincide either with (i) non-analytic changes in the eigenstates of the Hamiltonian governing the system (such
as the far higher melting temperature $T_{melt}>T_{0}$) or (ii) are associated with a singular temperature dependence of a probability distribution function that we introduce.
We will explain how the increasing relaxation times can appear naturally without a phase transition at any positive temperature $T \neq T_{melt}$. Our investigation will lead to a fit very different from that of VFTH that, in its minimal form, has only one parameter. More broadly, we will illustrate that {\it quantum} considerations for classical (i.e., non-cryogenic) liquids imply that upon supercooling, dramatically increasing relaxation times may appear without corresponding large thermodynamic signatures. To avoid confusion, we stress a simple maxim. Quantum mechanics may afford a practical {\it ``computational shortcut'' } to (semi-) classical calculations. Whenever, in the spirit of all earlier works, one may think about non-cryogenic liquids in strictly classical terms then a ``quantum'' state $| \psi \rangle$ can be viewed as merely a crutch to facilitate the analysis of the system evolution. We briefly expand on this statement. Given a time independent Hamiltonian $H$, the classical many body system evolves under Hamilton's equations of motion. With all observables promoted to operators, any wavefunction $| \psi \rangle$ (whether a real physical quantum state or a fictive classical crutch) that yields the classical phase space coordinates ($\langle \psi| x_{\alpha}| \psi  \rangle, ~ \langle \psi| p_{\alpha}| \psi  \rangle$) at time $t$, will (when the uncertainties in particle locations are small \cite{ehrenfest} or may be emulated by external thermal noise in the classical system) automatically, produce the classical phase space coordinates ($ \langle \psi| e^{iH \Delta t/\hbar} x_{\alpha} e^{-iH \Delta t/\hbar} | \psi  \rangle,  ~ \langle \psi| e^{iH \Delta t/\hbar}  p_{\alpha} e^{-iH \Delta t/\hbar} | \psi  \rangle$) at time $(t + \Delta t)$. Thus if, as we will see later on, due to simple {\it ``selection rules''}, certain features are transparent in the quantum evolution given by $e^{-iH \Delta t/\hbar}$ then these features may effortlessly lead to results that can be very difficult to directly derive via a conventional classical analysis (although they are, nevertheless, also a consequence of the classical equations of motion) \cite{e2+}. A quantum mechanical framework is, of course, fundamentally also more complete in other regards. The ``counting'' and/or weighing of classical ``microstates'' often implicitly requires underlying quantum notions. As is well appreciated, in addition to the Gibbs paradox, this is vividly seen when introducing Planck's constant $h$ in phase space integrals. For instance, (semi-) classically, the textbook partition function \cite{huang} of a system of $N$ distinguishable particles held at an inverse temperature $\beta=1/(k_{B} T)$ in $D$ spatial dimensions 
is $Z = h^{-DN} \int d^{DN} x ~d^{DN} p~ ~e^{-\beta H}$.
Similarly, Planck's constant appears when computing the entropy in the microcanonical ensemble (wherein each (semi-) classical ``microstate'' occupies a hypercube of volume $h^{DN}$ in phase space). The approach employed in the current work will enable us to infer the viable weight of each of the correct corresponding ``microstates''. As alluded to above and as we will 
expound, this weighing will be achieved via an eigenstate decomposition. 

\section{The disorder free many-body Hamiltonian.}
Unlike ``spin-glass'' systems \cite{SG,SG1} having quenched disorder, the supercooled liquids described above have no externally imposed randomness; these liquids could crystallize if cooled slowly enough. For emphasis, we write the {\it exact} many-body Hamiltonian of disorder free liquids,
\begin{eqnarray}
\label{atomic}
H= -\sum_{i} \frac{\hbar^{2}}{2M_{i}} \nabla^{2}_{R_{i}} - \sum_{j} \frac{\hbar^{2}}{2m_{e}} \nabla_{r_{j}}^{2} - \sum_{i,j} \frac{Z_{i}e^{2}}{|R_{i}-r_{j}|} \nonumber
\\  + \frac{1}{2} \sum_{i \neq i'} \frac{Z_{i} Z_{i'} e^{2}}{|R_{i} - R_{i'}|} + \frac{1}{2} \sum_{j\neq j'} \frac{e^{2}}{|r_{j} - r_{j'}|}.
\end{eqnarray}
In Eq. (\ref{atomic}), $M_{i}, R_{i},$ and $Z_{i}e$ are, correspondingly, the mass, position, and charge of the $i-$th nucleus, 
while $r_{j}$ is the location of the $j-$th electron (whose mass and charge are $m_{e}$ and $(-e)$ respectively). In systems of practical interest, the number of ions and electrons 
is very large. Finding tangible exact (or even approximate) eigenstates of the Hamiltonian of Eq. (\ref{atomic}) is impossible (or, at best, is extremely challenging). Insightful simplified variants of this Hamiltonian have been extremely fruitful in various areas of physics and chemistry. Our more modest goal is not to solve the spectral problem posed by Eq. (\ref{atomic}) nor to advance any intuitive approximations in various realizations. We will only rely on the mere {\em existence} of this Hamiltonian and that of its corresponding eigenstates. We will denote the eigenstates of $H$ by $\{|\phi_{n} \rangle\}$ and mark their corresponding energies by $\{E_{n}\}$. As mentioned to in the Introduction and will next be made evident in Section \ref{generalc}, different from all other approaches to date (e.g.,\cite{paw,kt,fragile1,fragile2,ag,ag1,ag2,ag3,CG,ejcg,elm,jam,jam',gt1,sri,gt2,gt3,gt4,gt5,gt6,moore,myega,wg,mdd,avoided1,avoided2,avoided3,vft1,vft2,vft3,vft4,vft5}) that assume the existence of numerous special temperature scales associated with glass formation, the only temperature that we will invoke is {\it the measured equilibrium melting transition temperature $T_{melt}$ of the system of Eq. (\ref{atomic}).} 

\begin{figure*}
\centering
\includegraphics[width=1.5 \columnwidth]{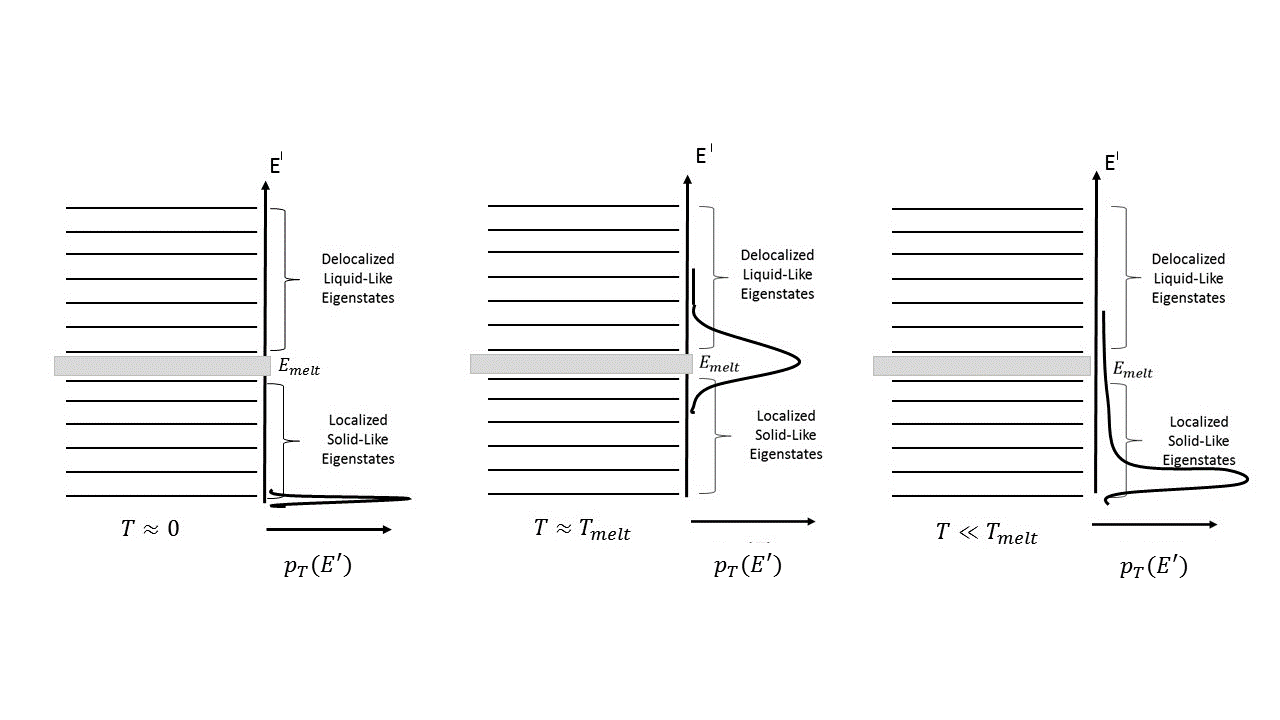}
\caption{A schematic description of the eigenstates of Eq. (\ref{atomic}) and possible probability distributions ${\it{p}}_{T}(E')$ (Eq. (\ref{pt+})) at different temperatures $T$. The standard thermodynamic equilibrium averages of Eqs. (\ref{mc}, \ref{etheq}) enabled us to establish that at energy densities below 
that of melting, the eigenstates are localized solid-like states while at energies above melting the eigenstates are delocalized fluid type states.
In the vicinity of the melting temperature $T_{melt}$, there is (in general) an interval of energies associated with {\it latent heat} (denoted 
in the text by ${\cal{PT}}E{\cal{I}}$). We sketch a Gaussian type probability distribution
${\it{p}}_{T}(E')$ at different temperatures. This probability distribution must satisfy Eq. (\ref{constraint}) and thus shifts to lower energies
as the temperature is lowered. At very low temperatures, only a small cumulative probability $\int_{E_{melt}}^{\infty}
{\it{p}}_{T}(E') dE'$ is associated with energies above the melting transition. Thus, at low $T$ the system may be nearly localized and exhibit extremely large relaxation times. }
\label{eigen.}
\end{figure*}

\section{The general character of the eigenstates at different energies.}
\label{generalc}
For our purposes, it will suffice to know the distinctive features of the eigenstates of Eq. (\ref{atomic}) which we will ``read off'' from knowledge of the thermodynamic behavior of equilibrated systems defined by Eq. (\ref{atomic}) at different temperatures. 
In the equilibrium microcanonical (mc) ensemble, the average of any operator ${\cal{O}}$ at an energy $E$ is
\begin{eqnarray}
\label{mc}
\langle {\cal{O}}(E) \rangle_{mc} \equiv \frac{1}{{\cal{N}}[E-\Delta E,E]}  \sum_{E - \Delta E \le E_n \le E}  \langle \phi_n| {\cal{O}}| \phi_n \rangle.
\end{eqnarray}
In Eq. (\ref{mc}), $\Delta E$ is a system size independent energy window and ${\cal{N}}[E-\Delta E, \Delta E]$ is the number of the eigenstates $|\phi_n \rangle$ 
of energy $E_{n}$ lying in the interval $E- \Delta E \le E_n \le E$. Typically, we demand that $\Delta E/E \to 0$ in the thermodynamic limit. By equating the lefthand side of Eq. (\ref{mc}) to the measured equilibrium value, this standard relation becomes a {\it dictionary} between the many body states (appearing on the righthand side of Eq. (\ref{mc})) and the measured quantities in equilibrium (equal to the average $\langle {\cal{O}}(E) \rangle_{mc}$). 
In the extreme case of small $\Delta E$, only a single eigenstate (or a set of degenerate eigenstates) lies in the interval $[E-\Delta E,E]$, i.e.,
$\langle {\cal{O}}(E) \rangle_{mc} \to
\langle  \phi_{n} | {\cal{O}}| \phi_{n} \rangle$
(or the average of such degenerate eigenstates). 
In such a case,
\begin{eqnarray}
\label{etheq}
Tr(\rho_{eq} {\cal{O}}) = \langle  \phi_{n} | {\cal{O}}| \phi_{n} \rangle.
\end{eqnarray}
In Eq. (\ref{etheq}), $\rho_{eq}$ is the equilibrium (micro-canonical) density matrix associated with energy $E=E_{n}$ or (when ensemble equivalence applies) the canonical density matrix associated with a temperature $T$ such that the internal energy $E(T) = E_{n}$. Within the canonical ensemble at temperature $T$, the density matrix $\rho_{eq} = \exp(-\beta H)/Z$ with $Z$ the partition function. 
When several eigenstates share the same energy $E_{n}$ then 
the right hand side of the above equation will be replaced by an average over degenerate eigenstates.
When Eq. (\ref{etheq}) holds, the system satisfies the ``Eigenstate Thermalization Hypothesis'' \cite{eth1,eth2,eth3,rigol,pol,polkovnikov1,polkovnikov2,von_neumann}. 
``Many body localized'' systems \cite{MBL1,MBL2,MBL3,MBL4,MBL5,MBL6} (that we will further discuss in Appendix \ref{mbl-a}), particularly those with disorder, do not thermalize and may violate Eq. (\ref{etheq}) even at infinite temperature. 

Our first, very simple, observation is that the disorder free Hamiltonian of Eq. (\ref{atomic}) is {\it experimentally} known to lead to an equilibrated solid at low temperatures and an equilibrated liquid (or gas) at higher temperatures or energies. In the low energy solid phase, the system may break rotational and translational symmetries and is not fully ergodic;
there are degenerate states $\{|\phi_n \rangle\}$ that are related to each other by such symmetry operations. Stated equivalently, these low energy
solid-like eigenstates need not transform as singlets under all symmetries of the Hamiltonian. For these low energy states, we may apply the microcanonical ensemble relation of Eq. (\ref{mc}) when the lefthand side is confined to a subvolume of phase space (associated with the quantum numbers defining $|\phi_n \rangle$) over which the system is ergodic and is thermally equilibrated. Eq. (\ref{mc}) and the more refined Eigenstate Thermalization equality of Eq. (\ref{etheq}) may then be invoked for computing thermodynamic observables whose behaviors are empirically known. Since the character of the averages on the lefthand sides of Eqs. (\ref{mc}, \ref{etheq}) is empirically known, we will be able to ascertain the nature of the expectation values on the righthand sides of Eqs. (\ref{mc}, \ref{etheq}). The equilibrium problem posed by Eq. (\ref{atomic}) leads to solids at low temperatures (or energy densities),
liquids at energies above melting, gases above the boiling temperatures, and so on. With $V$ the volume, it follows that, {\it{as a function of the energy density $(E_{n}/V)$, the eigenstates $|\phi_n \rangle$ of Eq. (\ref{atomic}) undergo corresponding transitions}}. Thus, 

{\bf (A)}  Eigenstates of Eq. (\ref{atomic}) of energy density larger than $(E_{melt}/V)$ are {\it delocalized liquid like states}.

{\bf (B)} Eigenstates of Eq. (\ref{atomic}) of energy density smaller than $(E_{melt}/V)$ are {\it localized solid like states}.

By the two qualifiers of {\it liquid} and {\it solid} like states, we simply refer to the fact that the averages of {\it any} observable ${\cal{O}}$ as computed by Eqs. (\ref{mc}, \ref{etheq}) will, correspondingly, lead to
the observed value of ${\cal{O}}$ in an equilibrated liquid or solid \cite{explain_quantum_number}. Setting this operator to be the Hamiltonian itself, ${\cal{O}} =H$, allows us to self-consistently infer that the energy density of the equilibrated system at the melting
temperature $(E_{melt}/V)$, sets the boundary between the liquid and solid like states. When the system volume $V$ is not held fixed, the energy per atom ($E_{melt}/N$) governs the transitions between the localized to delocalized states. In our analysis, we {\it do not require} that the low temperature equilibrated solid is a crystal. The sole assumption made is that at low temperatures (or energies), the equilibrated system forms a solid, whether crystalline or
of any other type. These equilibrated solids (and associated eigenstates) may exhibit phonon type excitations and other behaviors associated with standard solids. 
To illuminate certain aspects of the underlying physics, we will, at times, refer to crystalline systems. 
In Fig. \ref{eigen.}, we sketch this conclusion. 
 If the lowest eigenstates of Eq. (\ref{atomic}) correspond to a crystalline solid then as energy is increased
the first non-analyticity will appear at the melting energy, $E=E_{melt}$. Whenever a {\it latent heat of fusion} $ \Delta Q_{fusion}$ is absorbed/released during heating/cooling between the equilibrium solid and liquid phases at the melting temperature $T_{melt}$, then by Eqs. (\ref{mc}, \ref{etheq}), there must be a {\it range} of energy densities where the eigenstates of Eq. (\ref{atomic}) exhibit coexisting liquid and solid-type structures. Thus, more precise than the broad-brush statements of (A) and (B) above, the energy density bounds on the solid like and liquid states are set by $(E^{\pm}_{melt}/V)$ where $E_{melt}^{\pm}$ correspond to the upper and lower limits of this latent heat interval. In Fig. (\ref{eigen.}), we christen this liquid-solid coexistence region to be the ``phase transition energy interval''  (designated henceforth as ${\cal{PT}}E{\cal{I}}$) 
by the shaded region near $E_{melt}$. For simplicity, in later sections, we will denote the energy density of the lowest lying liquid like states
(i.e., those just above the ${\cal{PT}}E{\cal{I}}$) by $(E_{melt}/V)$ (i.e., we will denote $E_{melt}^{+}$ by $E_{melt}$). 
Below the coexistence region, at energies densities $E/V< E_{m}^{-}/V$, liquid type microstates (or eigenstates) may still exist yet their relative weight is exponentially smaller (in the system size)
than that of the solid-type states (their free energy density is higher than that of the solid type states). While we do not wish to further complicate our discussion, we remark, for completeness and precision, that in general glassformers, the above latent heat region does not correspond to a single melting temperature $T=T_{melt}$, but is rather bounded between the so-called ``liquidus'' temperature (above which the system is entirely liquid) and the ``solidus'' temperature (below which the equilibrium system is completely solid), $T_{liquidus} \ge T \ge T_{solidus}$. 
If there are additional crossover temperatures in the liquid then these imply corresponding 
crossovers in the eigenvectors at the associated energy densities. For instance, it has been found that liquids may fall out of 
equilibrium at temperatures below $T_A > T_{melt}$ (as evidenced by e.g., the breakdown of the Stokes-Einstein relation in metallic liquids \cite{numerics',egami}).
If this falling out of equilibrium is not a consequence of rapid cooling, it must be that the eigenstates of $H$ change character at the energy density associated
with $T_A$. Determining these and other crossover temperatures (if they are indeed there) is not an easy task. The equilibrated system exhibits flow (and thermodynamic features of the liquid) 
above the liquidus or melting temperature while it is completely solid below the solidus temperature and cannot flow at all. To avoid the use of multiple parameters and temperatures,
in the simple calculation and the viscosity fit that we will derive, we will largely assume that the eigenstates change character at the unambiguous well measured liquidus or melting temperature. 

 As we will emphasize to throughout the current work, the system is described by a probability density matrix $\rho$. We can obtain the probabilities of being in different pure states by diagonalizing this matrix. To make our ideas lucid in what will follow in this work and to avoid any extraneous complications, we will often analyze only a typical pertinent pure state of high probability as an example.
 
 Even though, barring many body localized states and special dependencies \cite{explain_quantum_number}, the eigenstates of energy densities lower than that of the equilibrium melting temperature, i.e., $E_{melt}/V$
(or the more stringent bound of $(E_{melt}^{-}/V)$) emulate, in the average sense defined by the microcanonical mean of Eq. (\ref{mc}), the behavior of the equilibrium solid, the supercooled system can nevertheless exhibit hydrodynamic flow at long times (i.e., the system viscosity is finite). This occurs since the state of the supercooled system differs from that of a simple eigenstate. In the next section, we turn to the state generated by supercooling. As we will describe, the state of the system after supercooling will, generally, {\it contain both liquid like and solid like eigenstates}. In particular, the existence of liquid like contributions even when the temperature (or, equivalently, the energy density) enables (slow) hydrodynamic flow in the supercooled liquid.

  %To outline our conceptual framework and simplify our expressions, we will largely omit the discussion of latent heat effects. 

\section{Supercooling as an evolution operator.}
\label{ScS}

Formally, an equilibrated liquid in an initial state 
$|\psi (t_{initial}) \rangle$ is
supercooled to a final state $| \psi \rangle$ at time $t=t_{final}$ via an evolution operator,
$| \psi \rangle = \tilde{U} (t_{final}, t_{initial}) | \psi (t_{initial}) \rangle$; ~
$\tilde{U}(t_{final}, t_{initial}) = {\cal{T}}  e^{-\frac{i}{\hbar} \int_{t_{initial}}^{t_{final}} dt' \tilde{H}(t')}$,
where ${\cal{T}}$ denotes time ordering. The probability density matrix evolves as $\rho(t_{final}) = \tilde{U}(t_{final}, t_{initial}) \rho(t_{initial}) \tilde{U}^{\dagger}(t_{final}, t_{initial})$. We will assume a particular idealized cooling procedure. 
A time dependent Hamiltonian $\tilde{H}(t) \neq H$ will replace
the Hamiltonian of Eq. (\ref{atomic}) during the time interval $t_{initial} \le  t \le  t_{final}$. During those times, the Hamiltonian $\tilde{H}(t)$ will include coupling to external sources (e.g., photons or chassis phonons) that lead to the supercooling. These external sources may, in some cases, be emulated by, e.g., Caldeira-Leggett type and similar couplings wherein the environment is represented by harmonic oscillators
that couple bilinearly  \cite{C-L} to the generalized coordinates of the liquid. Such a coupling leads to an effective Hamiltonian $\tilde{H}(t)$ once
the bath degrees of freedom are integrated out. At all other times (i.e., $t < t_{initial}$ or $t > t_{final}$), we will assume that the Hamiltonian $\tilde{H}(t)$ is given by $H$ of Eq. (\ref{atomic}),
\begin{eqnarray}
\label{HHH}
\tilde{H}(t<t_{initial}) =H  &\to& \tilde{H}(t_{initial} \le t \le t_{final}) \neq H \nonumber
\\  &\to& \tilde{H}(t >t_{final}) =H.
\end{eqnarray}
In order to allow for a change in the energy $\langle H \rangle$, the commutator $[\tilde{U}(t_{final}, t_{initial}), H] \neq 0$ \cite{rejuv}. If the temperature (or energy) of the system does not vary (or is nearly fixed) at long times then the time derivative 
$0=d/dt (\langle \psi (t) | H | \psi (t) \rangle) = \frac{i}{\hbar} \langle \psi(t) |[\tilde{H}, H] | \psi(t) \rangle$. Thus, if for times $t>t_{final}$, the energy $\langle \psi(t)| H | \psi(t) \rangle$ is time independent then $ \langle \psi (t)|  [\tilde{H},H]  | \psi (t) \rangle =0$. Consequently, for times $t>t_{final}$, the Hamiltonian $\tilde{H}$ may generally be that of Eq. (\ref{atomic}) with no additional nontrivial terms. This is indeed realized in  Eq. (\ref{HHH}). If the energy as measured by $H$ is very slowly varying with time for $t>t_{final}$, then we may still employ Eq. (\ref{HHH}) with negligible corrections.

 Regardless of specific model Hamiltonians $\tilde{H}(t)$ emulating particular cooling protocols, after supercooling (i.e., at time $t=t_{final}$), we may expand $| \psi \rangle$ in the complete eigenbasis of the Hamiltonian $H$ of Eq. (\ref{atomic}),
\begin{eqnarray}
\label{dec}
|\psi \rangle= \sum_{n} c_{n} | \phi_{n} \rangle.
\end{eqnarray} 
When Eq. (\ref{etheq}) holds, an eigenstate of $H$ corresponds to an equilibrated system. Similar conclusions may be drawn, via the microcanonical ensemble average of Eq. (\ref{mc}), for a superposition of eigenstates over a narrow energy interval $\Delta E$. The {\it adiabatic theorem} of quantum mechanics (e.g., \cite{cite_adiabatic}) implies that if $|\psi (t_{initial}) \rangle$ is an eigenstate of the (initial Hamiltonian) $H$ then
 for slowly varying $\tilde{H}(t)$, the final state $| \psi \rangle$ is also an eigenstate of the (final Hamiltonian) $H$. However, in the {\it diametrically opposite} limit of a non-adiabatic $\tilde{H}(t)$ embodying the supercooling,
 the final state $| \psi \rangle$ is, generally, not an eigenstate of $H$. A typical initial thermal state is a superposition of eigenstates of $H$ of (nearly) the same energy. After the evolution with ${\tilde{U}}(t_{final}, t_{initial})$, this state will become a linear sum of eigenstates of $H$ of largely varying energies. Indeed, as the supercooled liquid is, by its nature, out of equilibrium, a {\it finite range} of energy densities must appear in the sum of Eq. (\ref{dec}). (We will further elaborate on this in Section \ref{sec:prob}.) The feature that is of crucial importance is that the probability density, 
\begin{eqnarray}
\label{pt+}
{\it{p_{T}}}(E') \equiv \sum_{n} |c_{n}|^{2} \delta(E'-E_{n}),
\end{eqnarray} 
may have contributions from both (i) low energy solid-like states ($E_{n} < E_{melt}$) and (ii) higher energy fluid-type eigenstates 
($E_{n} > E_{melt}$). (Eq. (\ref{pt+}) may, be trivially extended from a single pure state to the density matrix describing a general open system (as in, e.g., Eq. (\ref{ptttt}) that we will turn to later)). 

The idealized protocol of Eq. (\ref{HHH}) leads, after to supercooling (i.e., at times $t > t_{final}$), to the time independent probability density ${\it{p}}_{T}$ of Eq. (\ref{pt+}) for the closed system
defined by the Hamiltonian $H$. In supercooled liquids of experimental interest,
coupling to the environment will not render the probability density for having an energy $E'$ to be time independent at times $t> t_{final}$. 
In particular, at long enough times $t$, the system may return to its ideal equilibrium state and have a probability density ${\it{p_{T}}}$ that is delta-function like in the energy density (and for which the narrow energy density canonical average or microcanonical average of Eq. (\ref{mc}) apply). However, the time required to achieve such a true equilibrium state may be far larger than feasible experimentally. Indeed, if a liquid is supercooled below melting ($T< T_{melt}$) it will not, on relevant experimental time scales, exhibit structural or other properties of an equilibrated solid or crystal. (Therefore, as we will return to, on these time scales, the distribution ${\it{p}}_{T}$ associated with the supercooled system must be different from that defining an equilibrium system.) The above ``crystallization'' time scale (after which the system
emulates a true equilibrium solid such as a crystal having a delta-function distribution in the energy density) differs from the relaxation time of the supercooled liquid to its long time state following a perturbation.  

In the equations that follow, we will largely use a prime superscript to denote quantities when these are evaluated for the equilibrated thermal system associated with the Hamiltonian of Eq. (\ref{atomic}). As the temperature $T$ to which the liquid is supercooled corresponds to an internal energy $E_{final} = E(T)$ of the equilibrated system, we have 
%the first moment of the distribution ${\it{p_{T}}}$ is set by
\begin{eqnarray}
\label{constraint}
 E(T)  =  \int dE'~ {\it{p_{T}}}(E')E' \equiv \langle E \rangle.
\end{eqnarray}
The temperature of the non-equilibrium supercooled liquid may be measured, e.g., by pyrometry. The emitted photons probe the average effective temperature of the supercooled liquid. 
Thus, even though the supercooled system is out of equilibrium (and exhibits fascinating memory and aging effects) 
and the notion of temperature is somewhat subtle, the energy as given by Eq. (\ref{constraint}) is well defined. 
At sufficiently high temperatures $T_{high}>T_{melt}$, the experimentally measured specific heat of the supercooled liquid is equal to that of the equilibrated liquid. This implies that, up to experimental accuracy, at high temperatures $T_{high}$, the internal energies of the supercooled liquid and the annealed liquid may be set to be the same, $E(T_{high}) = E'(T_{high})$.  The distribution ${\it{p_{T}}} (E')$ must satisfy the condition that the associated energy $E$ may be computed from the measured heat capacity $C_{v}$,
\begin{eqnarray}
\label{ECE}
E(T) = E(T_{high}) + \int _{T_{high}}^{T} dT^*~ C_{v}(T^*) = \int dE'~ {\it{p_{T}}}(E')E'.
\end{eqnarray}
The second equality in Eq. (\ref{ECE}) reiterates that of Eq. (\ref{constraint}). 
Most conventional theories of glasses do not allow for the dependence of measurable dynamic and thermodynamic quantities on the history of the preparation of the glass. In sharp contrast, our approach naturally {\it allows for such a history dependence}. 
For different cooling and reheating-recooling (``rejuvenation'') protocols and/or those 
involving different perturbations embodying external shear, etc., the evolution $\tilde{U}(t_{initial}, t_{final})$ and consequently the 
distribution ${\it{p_{T}}}(E')$ will be different from those associated with the standard supercooling procedures. 

At all times after supercooling, $t>t_{final}$ (Eq. (\ref{HHH})), the Hamiltonian becomes again that of Eq. (\ref{atomic}) and $| \psi(t) \rangle = e^{-iH(t-t_{final})/\hbar} | \psi \rangle$. As $| \psi \rangle$ is not an eigenstate of $H$, the system may evolve nontrivially with time. Nonetheless, the long time averages may simplify as we will explain in Section \ref{lta-sec}. Before doing so, we briefly touch on another aspect.

\section{Long time averages of local observables in the supercooled state.}
\label{lta-sec}
Given Eq. (\ref{dec}), we next perform standard quantum mechanical calculations similar to those appearing in works on the Eigenstate Thermalization Hypothesis \cite{eth1,eth2,eth3,rigol,pol,polkovnikov1,polkovnikov2}. 
The long time average (l.t.a.) of a quantity ${\cal{O}}$ which we may consider to be a local operator (or derived from a sum of such operators),
\begin{eqnarray}
\label{llong}
{\cal{O}}_{l.t.a.} = 
\lim_{ \tilde{{\mathcal{T}}} \to \infty}
\frac{1}{\tilde{\mathcal{T}}}
 \int_{t_{final}}^{t_{final} + \tilde{{\mathcal{T}}}} dt'~  \langle \psi (t') | {\cal{O}} | \psi (t') \rangle \nonumber
 \\ =
\lim_{ \tilde{{\mathcal{T}}} \to \infty}
\frac{1}{\tilde{\mathcal{T}}}
\sum_{n,m} c_{n}^{*} c_{m} \langle \phi_n| {\cal{O}} | \phi_m \rangle \nonumber
\\ \times  \int_{t_{final}}^{t_{final} + \tilde{{\mathcal{T}}}}  dt'~ e^{i(E_{n}-E_{m})(t'-t_{final})/\hbar}.
\end{eqnarray}
The long time average of $\lim_{{\tilde{\mathcal{T}}} \to \infty} \frac{1}{{\tilde{\mathcal{T}}}} \int_{0}^{
{\tilde{\mathcal{T}}}} dt' e^{i (E_{n}- E_{m}) t'/\hbar}$ vanishes if $E_n \neq E_m$. Thus, barring special commensuration, in the long time limit of Eq. (\ref{llong}), only (i) diagonal matrix elements of ${\cal{O}}$ and (ii) matrix elements of ${\cal{O}}$ between degenerate states will remain. 
We will shortly explicitly turn (in Eqs. (\ref{lllong},\ref{shorteq},\ref{richer_dep})) to the situations in which degenerate states appear. 
Properties (i) and (ii) enable us to relate the standard long time average of Eq. (\ref{llong}) computed with the Hamiltonian of Eq. (\ref{atomic}) (for which microcanonical averages of general observables ${\cal{O}}$ are equal their to their measured equilibrium values) to
a simple new expression,
\begin{eqnarray}
\label{thermog}
 {\cal{O}}_{l.t.a.; s.c.} &&= \int_{0}^{\infty} dE'~  {\it{p_{T}}} (E') ~ {\cal{O}} (E') \nonumber
\\ &&=  \int_{0}^{\infty} dT' ~{\cal{O}}(E'(T')) ~{\it{p}}_{T} \big(E'(T')\big) ~ C'_{V}(T') \nonumber
\\ &&+ \int_{{\cal{PT}}E{\cal{I}}} dE' ~ {\it{p_{T}}} (E') ~ {\cal{O}} (E').
\end{eqnarray}
In Eq. (\ref{thermog}), we employed the shorthand 
${\cal{O}}(E') \equiv \langle {\cal{O}}(E') \rangle_{mc}$
for the {\it equilibrium} microcanonical average of ${\cal{O}}$ at a fixed energy $E'$ (Eq. (\ref{mc})). Thus,
this equation transforms the long time average of a general operator ${\cal{O}}$ in the supercooled state into a weighted integral over the averages of ${\cal{O}}$ in an equilibrated solid or liquid at
energies $E'$ (and respective temperatures $T'$) . 
Since the Hamiltonian of Eq. (\ref{atomic}) is empirically known to equilibrate,
the microcanonical average ${\cal{O}}(E')$ is indeed equal to the measured value of
${\cal{O}}$ at for a system energy $E'$ (justifying our abbreviated notation of ``${\cal{O}}(E')$'').
The principal implicit assumption invoked to obtain our relation of Eq. (\ref{thermog}) concerns the replacement of the diagonal matrix elements $\langle \phi_n| {\cal{O}} | \phi_n \rangle$ in Eq. (\ref{llong}) by their average value over a narrow energy shell centered about $E'$ (where $E'= E_{n}$). Such an assumption is valid if $\langle \phi_n| {\cal{O}} | \phi_n \rangle$ is a regular function of the energy $E_{n}$. As we discussed earlier, we anticipate such a regular behavior if no equilibrium phase transitions appear at the energy $E_{n}$. In Appendix \ref{mbl-a}, we will further comment on this regularity assumption. Colloquially, Eq. (\ref{thermog}) asserts that the long time average is intrinsically {\it classical}: the operator ${\cal{O}}$ can simply be replaced by its c-number values in different states; it embodies a probabilistic classical average (in which the averaged over classical ``microstates'' are chosen to be the eigenstates of the Hamiltonian). This equality realizes the classical correspondence maxim outlined in the Introduction. We note that, in the classical (i.e., ``$\hbar \to 0$'') limit, the right-hand side of Eq. (\ref{llong}) approaches Eq. (\ref{thermog}) also for arbitrarily small ${ \tilde{{\mathcal{T}}}}$; there is, indeed, no remnant of finite $\hbar$ quantum mechanics in our relation of Eq. (\ref{thermog}). Thus, although our analysis is quantum and we invoke the spectral decomposition of the state $| \psi \rangle$, formally, our calculations yield classical results when choosing the summed over ``microstates'' to be the eigenstates of the Hamiltonian. In the last equality of Eq. (\ref{thermog}), the integrand of the first term is evaluated for an equilibrated system at a temperature $T' (\neq T_{melt})$; ~ $E'$ is the internal energy of the equilibrated system at a temperature $T'$ and $C'_{V}(T') = (dE'/dT')$ is the heat capacity at constant volume.
We recall that a range of possible equilibrium energies $E'$ spans the ${\cal{PT}}E{\cal{I}}$ latent heat interval (in which the temperature $T'$
of the equilibrium system is fixed, $T^{'}=T_{melt}$); this energy range is captured in the last integral of Eq. (\ref{thermog}). The empirical value of the energy $E$ at all temperatures may be obtained by integrating the heat capacity (Eq. (\ref{ECE})). Although (as $\hbar \neq 0$), the result of Eq. (\ref{thermog}) strictly holds only in the asymptotic long time limit, in reality {\it a very rapid convergence} of the off-diagonal ($n \neq m$) terms (of different energies) occurs in many cases \cite{rapid} and already at very short time intervals, ${ \tilde{{\mathcal{T}}}}$, the off-diagonal contributions in 
Eq. (\ref{llong}) may vanish in systems that thermalize and the long time quantum average becomes classical.

In our key Eq. (\ref{thermog}), the distribution ${\it{p}}_{T}$ is the only quantity not known from experimental measurements of the equilibrated system. By Eq. (\ref{thermog}), 
if the range of energies over which an assumed analytic ${\it{p}}_{T}(E')$ has its pertinent support
does not correspond to an energy density (or associated $T'$ such that the energy is that of the equilibrium internal energy at that temperature, $E'= E'(T')$) at which the equilibrated system exhibits a phase transition then, all observables ${\cal{O}}$ will not display the standard hallmarks of a phase transition. Thus, we now substantiated the claim made at the beginning of the current work. Namely, for equilibrium states of the Hamiltonian of Eq. (\ref{atomic}), the lowest temperatures at which transitions may be discerned by probes ${\cal{O}}$ must either (1) correspond to the conventional melting or freezing temperature or, less plausibly, (2) are associated with a singular temperature dependence of the distribution ${\it{p}}_{T}(E')$ generated by supercooling. 
Eq. (\ref{thermog}) is quite powerful. Features (and predictions) associated with ${\it{p}}_{T}(E')$ may be examined by using judicious single-valued measurable functions ${\cal{O}}(E)$. When the operators ${\cal{O}}$ are functions of the Hamiltonian, ${\cal{O}} = g(H)$ and are diagonal in the basis of the eigenstates of the Hamiltonian, Eq. (\ref{thermog}) becomes an identity that does not require a long time average. Thus, regardless of the complexity of the many body supercooled state, all expectation values of thermodynamic functions that may be derived from the internal energy are exactly given by Eq. (\ref{thermog}) and its derivatives.  
When probed by such general observables ${\cal{O}}$, phase transition singularities of the equilibrated system will, due to the finite width of the distribution ${\it{p}}_{T}(E')$, now be smeared over a finite temperature range. Thus, as a function of temperature, changes in such observables in the supercooled fluid (and their $T$ derivatives such as the specific heat) may exhibit crossovers at the melting temperature (instead of sharp changes as they do in the equilibrated system). This observation concerning smearing also applies to measures of structural order (such as density $\langle {\sf n}(\vec{x}) \rangle$) further discussed in Appendix \ref{space.}.

For completeness, we now briefly discuss the non-idealized case in which we are not confined to a single Schr\"odinger state $| \psi \rangle$ but rather employ the full density matrix $\rho \equiv \rho(t_{final})$ for an open system wherein different pure Schr\"odinger states $| \psi_{h} \rangle$ may appear with disparate probabilities $p_{h}$; in its diagonal form, the density matrix $\rho = \sum_{h} p_{h} | \psi_{h} \rangle \langle \psi_{h}|$. We will now also further explore the physical situation when degeneracies (such as those associated with translational and rotational symmetries of the Hamiltonian of
Eq. (\ref{atomic})) are present. Under these circumstances, we will have, in full generality,
\begin{eqnarray}
\label{lllong}
&&{\cal{O}}_{l.t.a.} = \nonumber
\\ && \lim_{ \tilde{{\mathcal{T}}} \to \infty}
\frac{1}{\tilde{\mathcal{T}}}
 \int_{t_{final}}^{t_{final} + \tilde{{\mathcal{T}}}} dt'~  Tr (e^{-iH(t'-t_{final})/\hbar} \rho e^{iH(t'-t_{final})/\hbar} {\cal{O}} ) \nonumber
 \\ =&& \sum_{n,m} \rho_{nm} {\cal{O}}_{mn} 
\lim_{ \tilde{{\mathcal{T}}} \to \infty}
\frac{1}{\tilde{\mathcal{T}}} \int_{t_{final}}^{t_{final} + \tilde{{\mathcal{T}}}}  dt'~ e^{-i(E_{n}-E_{m})(t'-t_{final})/\hbar}.
\end{eqnarray}
Here, as in Eq. (\ref{llong}), the long time average of the integral vanishes unless $E_{n} = E_{m}$. 
Whenever finite off-diagonal elements of ${\cal{O}}$ (i.e., the operator ${\cal{O}}_{mn} = \langle \phi_{m} | {\cal{O}}| \phi_{n} \rangle$ with 
$m \neq n$) appear between degenerate states ($E_{n} = E_{m}$),
we will diagonalize ${\cal{O}}$ in each such degenerate projected subspace of constant energy. We still employ the index $n$ to label the eigenstates
that follow this diagonalization of both (a) the Hamiltonian $H$ and (b) the operator ${\cal{O}}$ projected onto a space of constant energy, ${\cal{O}}_{P_{E'}} \equiv [\delta (E'-H) ~{\cal{O}}
~\delta (E'-H)]$. In the common eigenbasis of those two operators, Eq. (\ref{lllong}) reads
\begin{eqnarray}
\label{shorteq}
{\cal{O}}_{l.t.a.} =  \sum_{n} \rho_{nn} {\cal{O}}_{nn} \equiv Tr(\rho_{diag} {\cal{O}}).
\end{eqnarray}
Here, $\rho_{diag}$ is a diagonal matrix whose elements are $\rho_{nn}$.
Penetrating earlier works introduced, and discussed in depth, the diagonal density matrix $\rho_{diag}$ and its associated ``diagonal ensemble'' \cite{eth3,rigol,pol,polkovnikov1,polkovnikov2}. Eq. (\ref{shorteq}) constitutes a very simple extension of common earlier analysis to degenerate systems. Analogous to our transition from the standard equality of Eq. (\ref{llong}) to our new key relation of Eq. (\ref{thermog}), we observe that if ${\cal{O}}_{nn}$ is a regular function of the energy $E_{n}$ (and of any other pertinent quantum numbers associated with degeneracies that we will comment on below), then we may replace the weighted sum over the matrix elements ${\cal{O}}_{nn}$ in Eq. (\ref{shorteq}) by a
corresponding average over the equilibrium microcanonical averages of the observable ${\cal{O}}$ at an energy $E_{n}$ (and of any other relevant
quantum numbers). Thus, if the product $ \rho_{nn} {\cal{O}}_{nn}$ depends solely on the energy $E'$, then
Eq. (\ref{shorteq}) will imply Eq. (\ref{thermog}) with the identification
$\rho_{nn} \to {\it{p}}_{T}(E')$ (where $E^{'}=E_{n}$). We stress that, in comparing Eq. (\ref{thermog}) with Eq. (\ref{shorteq}), 
the quantity $ {\it{p_{T}}} (E')$ is the diagonal entry
of the density matrix $\rho$ of the system at a temperature $T$ when this density matrix is written in the eigenbasis of the Hamiltonian (and simultaneous eigenbasis of the operator ${\cal{O}}$ in the projected space of constant energy $E'$ in those cases in which the matrix elements of ${\cal{O}}$ do not vanish between degenerate energy eigenstates). 
An immediate corollary of Eqs. (\ref{thermog},\ref{lllong},\ref{shorteq}) (readily demonstrated by substituting ${\cal{O}} = (W-W_{l.t.a.})^2$ in these equations) that we will return to later on is that the variance, $\sigma_{W}^{2}$, of any quantity $W$ when it is computed with the diagonal density matrix $\rho_{diag}$ or, equivalently, 
when calculated with the distribution ${\it{p}}_{T}$, is equal to the square of the long time average of the squared fluctuations of $W$, i.e., 
$\sigma_{W}^{2} = \lim_{ \tilde{{\mathcal{T}}} \to \infty} \frac{1}{{\tilde{\mathcal{T}}}}
\int_{t_{final}}^{t_{final} + \tilde{{\mathcal{T}}}} dt' ~(W(t') - W_{l.t.a.})^2$. Since $\rho_{diag}$ is diagonal in the energy eigenbasis, this density matrix will, of course, not evolve with time, i.e., 
$\rho_{diag}(t') \equiv e^{-iH(t'-t_{final})/\hbar} \rho_{diag} e^{iH(t'-t_{final})/\hbar} = \rho_{diag}$. Such a time independence of the diagonal density matrix 
$\rho_{diag}$ is mandated by its role in the general (time-independent) long time average of Eq. (\ref{shorteq}). Thus, even if the full density matrix $\rho(t')$ evolves with time, $\rho_{diag}$ will still trivially be a function of the energy labeling the diagonal elements and, therefore, Eq. (\ref{thermog}) will remain valid. In realistic systems, external time dependent perturbations may further amend the Hamiltonian at times $t'>t_{final}$. When a general Hamiltonian $H(t')$
is not time independent, the function 
\begin{eqnarray}
\label{ptttt}
{\it{p}}_{T}(E') \equiv Tr [\delta(E'-H(t')) \rho(t') ],
\end{eqnarray}
a generalization of Eq. (\ref{pt+}),
will depend on the time $t'$. To aid analysis, in our case, 
we will invoke the idealized supercooling protocol of Eq. (\ref{HHH}) and thus, essentially, presume that the effect 
of fluctuations augmenting the unperturbed Hamiltonian  $H$ of Eq. (\ref{atomic}) at times $t'>t_{final}$ is slow relative to the time scale of the measurement of the observable ${\cal{O}}$.

We now expand on the situations wherein (1) both the Hamiltonian $H$ and the observable ${\cal{O}}$
(when the latter is projected to subspaces of constant energy (i.e., the projected operator ${\cal{O}}_{P_{E'}}$) commute
with an operator $\hat{{\cal{Q}}}$) 
and when (2) the diagonal matrix elements ${\cal{O}}_{nn}$ in the common eigenspace of the Hamiltonian $H$ and the projected
operator ${\cal{O}}_{P_{E'}}$ depend nontrivially on the eigenvalues this operator (which we will refer to as ``quantum numbers'' ${\cal{Q}}$). When a dependence on such additional quantum numbers other than the energy exists, Eq. (\ref{shorteq}) implies a sum or integration over all eigenvalues 
(or classical ``ergodic sectors'' \cite{nigel}) ${\cal{Q}}$. This leads to a trivial extension of Eq. (\ref{thermog}),
\begin{eqnarray}
\label{richer_dep}
 {\cal{O}}_{l.t.a.} &&= \int_{0}^{\infty} dE'~   \sum_{\cal{Q}}~ {\it{p_{T}}} (E'; {\cal{Q}}) ~ {\cal{O}} (E'; {\cal{Q}}) \nonumber
\\ &&= \int_{0}^{\infty} dT' ~{ \sum_{\cal{Q}}~ \cal{O}}(E'(T'),{\cal{Q}}) ~{\it{p}}_{T} \big(E'(T');{\cal{Q}} \big) ~ C'_{V}(T',{\cal{Q}}) \nonumber
\\ &&+ \int_{{\cal{PT}}E{\cal{I}}} dE' ~  \sum_{{\cal{Q}}} ~{\it{p_{T}}} (E';{\cal{Q}}) ~ {\cal{O}} (E',{\cal{Q}}). 
\end{eqnarray}
Here, ${\it{p_{T}}} (E';{\cal{Q}})$ is an abstraction of Eq. (\ref{pt+}) that includes the quantum numbers ${\cal{Q}}$. Namely,
\begin{eqnarray}
\label{ptq}
{\it{p_{T}}} (E';{\cal{Q}}) = Tr \Big(\rho ~\delta(E^{'}-H) ~\delta({\cal{Q}}- {\hat{{\cal{Q}}}})\Big).
\end{eqnarray}
We stress that ${\cal{Q}}$ labels the spectrum of ${\cal{O}}_{P_{E'}}$, whenever ${\cal{O}}_{P_{E'}}$ exhibits more than a single eigenvalue in a sector of fixed energy $E'$. Thus, in the general expressions of Eqs. (\ref{richer_dep},\ref{ptq}), the quantum number
${\cal{Q}}$ may implicitly depend on $E'$. 

For the observable ${\cal{O}}$ that will be of most interest to us in the current work (the terminal velocity that we will next discuss in Section \ref{vlv}), the best documented empirical dependence of the {\it equilibrium} expectation value ${\cal{O}}(E')$ is indeed that on the energy $E'$ (or equilibrium temperature) alone. That is, ${\cal{O}}_{P_{E'}}$ obtains a unique value in any sector of fixed energy $E'$. 
If, for this and other observables, no dependence on additional ergodic sector indices is experimentally indeed seen, then
we may employ Eq. (\ref{thermog}) instead of the more general relation of Eq. (\ref{richer_dep}) (with the identification ${\it{p}}_{T}(E') = \sum_{\cal{Q}}  {\it{p_{T}}} (E';{\cal{Q}})$). We will return to the richer possibilities afforded by Eq. (\ref{richer_dep}) \cite{note_loc} and discuss aspects that may augment straightforward, energy density, characteristics of viable many body localized states (Appendix \ref{mbl-a}). As we noted above, $\rho_{diag}$ is, identically, time independent even if $\rho(t')$ is not. That is,
the universal appearance of the time independent ${\it{p}}_{T}(E')$ (in any system with a time independent Hamiltonian) does not, of course, 
imply that the density matrix $\rho(t')$ is fixed. Indeed, a time independent density matrix $\rho$ typically only appears when the density matrix is a function of the energy and other ``constants of motion'',  $\rho = f(E, {\cal{Q}})$ while, as we underscored above, the time independence of ${\it{p}}_{T}(E')$ is identically guaranteed for any time independent Hamiltonian $H$. 

Our formalism may be broadened to allow for fluctuations in quantities other than the energy or particular quantum numbers.
The diagonal density matrix $\rho_{diag}$ may, for instance, generically be written in the basis 
in which the system volume is not held fixed (corresponding to the physical situation in which most experimental measurements are performed at constant pressure- not that of constant volume). If ${\cal{O}}$ is set to be the spatially averaged {\it particle density operator} ${\sf n}$, then a trivial extension of Eq. (\ref{thermog}) states that the long time density in the supercooled liquid ${\sf n}_{s.c.}$ can be computed from that of the number density ${\sf n}'(E',V')$ in the equilibrated solid at an energy $E'$ and volume $V'$. That is, rather explicitly, 
\begin{eqnarray}
\label{nET}
{\sf n}_{l.t.a.}(T,P) = \int dE' \int dV' ~  {\it{p_{T,P}}} (E',V') ~ {\sf n}' (E',V').
\end{eqnarray}
Here, the probability distribution ${\it{p_{T,P}}} (E',V')$ is an extension of ${\it{p}}_{T}(E')$ that now allows also for general fluctuations in the system volume $V'$ when the external experimentally measured pressure is $P$. Relations such as the above may afford consistency checks on the distribution ${\it{p}}$.  One may further trivially extend Eqs. (\ref{richer_dep}, \ref{nET}) to include long time averages of general observables ${\cal{O}}$ in which the volume and other parameters may change, i.e., 
\begin{eqnarray}
\label{o-average}
{\cal{O}}_{l.t.a.}&& =  \int dE' \int  \prod_{\alpha =1} ^{s} d \Lambda'_{\alpha}~ \sum_{\cal{Q}}  ~ {\it{p}}_{T, \lambda'_{1}, \cdots, \lambda'_{s}}(E', \Lambda_{1}, \cdots, \Lambda_{s}; {\cal{Q}})  \nonumber
\\ &&~ \times {\cal{O}} (E',\Lambda'_{1}, \cdots, \Lambda'_{s}; {\cal{Q}})  \nonumber
\\ &&=  \int_{0}^{\infty} dT'  ~ \int  \prod_{\alpha =1} ^{s} d \Lambda'_{\alpha} ~ \sum_{\cal{Q}} ~ {\it{p}}_{T, \lambda'_{1}, \cdots, \lambda'_{s}}(E', \Lambda_{1}, \cdots, \Lambda_{s}; {\cal{Q}})  \nonumber
\\ &&~ \times {\cal{O}} (E',\Lambda'_{1}, \cdots, \Lambda'_{s}; {\cal{Q}}) \nonumber
\\ &&+ \int_{{\cal{PT}}E{\cal{I}}} dE' ~   \int  \prod_{\alpha =1} ^{s} d \Lambda'_{\alpha}~ \sum_{{\cal{Q}}}  ~   {\it{p}}_{T, \lambda'_{1}, \cdots, \lambda'_{s}}(E', \Lambda_{1}, \cdots, \Lambda_{s}; {\cal{Q}})  \nonumber
\\ &&~ \times {\cal{O}} (E',\Lambda'_{1}, \cdots, \Lambda'_{s}; {\cal{Q}}).
\end{eqnarray}

In Eq. (\ref{o-average}), the probability distribution $ {\it{p}}_{T, \lambda_{1}, \cdots, \lambda_{s}}(E', \Lambda'_{1}, \cdots, \Lambda'_{s}; {\cal{Q}})$
depends on the energy $E'$, as well as any additional parameters $\{\Lambda'_{\alpha} \}_{\alpha=1}^{s}$ (e.g., the volume in Eq. (\ref{nET}) or other extensive variables) 
that define the system and any further quantum numbers ${\cal{Q}}$ (i.e., in this general case, eigenvalues of the operator ${\cal{O}}_{P_{E', \{\Lambda'_{\alpha}\}_{\alpha=1}^{s}}} \equiv [\delta (E'-H)
\prod_{\alpha=1}^{s} \delta(\Lambda'_{\alpha} - \Lambda_{\alpha}) 
 ~{\cal{O}} ~ 
\prod_{\alpha=1}^{s} \delta(\Lambda'_{\alpha} - \Lambda_{\alpha}) 
~\delta (E'-H)
]$). In the subscript of the probability distribution ${\it{p}}$ in Eq. (\ref{o-average}), along with the temperature $T$, the parameters $\{\lambda_{\alpha} \}_{\alpha=1}^{s}$ are the (intensive) quantities that are thermodynamically conjugate to the extensive variables $\{\Lambda_{\alpha}\}_{\alpha=1}^{s}$. Similar to the other equations that we derived, ${\cal{O}} (E',\Lambda'_{1}, \cdots, \Lambda'_{s}; {\cal{Q}})$
is the value of ${\cal{O}}$ in a microcanonical ensemble specified by $(E',\Lambda'_{1}, \cdots, \Lambda'_{s}; {\cal{Q}})$. Similar to the discussion following 
Eq. (\ref{ptq}), the quantum number ${\cal{Q}}$ may implicitly depend on the numbers $(E',\Lambda'_{1}, \cdots, \Lambda'_{s})$ specifying the projected operator ${\cal{O}}_{P_{E', \{\Lambda'_{\alpha}\}_{\alpha=1}^{s}}}$. Eq. (\ref{o-average}) may be applied to any system. In deriving Eq. (\ref{o-average}), we have merely used simple mathematical identities and the definition of the equilibrium microcanonical ensemble average for ${\cal{O}}$ over a states set by $(E',\Lambda'_{1}, \cdots, \Lambda'_{s};{\cal{Q}})$.
In what follows, we will largely employ the simplest variant of our relations- that of Eq. (\ref{thermog}).

\section{Relating the viscosity to a long time velocity average.} 
\label{vlv} 
In this and in the next section, we turn to dynamics and relaxation rates governing the viscosity. Traditionally, the viscosity of supercooled liquids is experimentally measured by finding the terminal velocity $v_{\infty, s.c.}$ of a dropping sphere. 
By Stokes' law for a low Reynolds number fluids (such as a viscous supercooled fluid), a sphere of radius $R$ dropped into a viscous fluid reaches a terminal velocity $v_{\infty; s.c.}= v_{l.t.a.; s.c.}$ set by the gravitational acceleration $g$, the viscosity $\eta$, and the mass densities $\rho_{sphere}$ and $\rho_{fluid}$ of the sphere and fluid respectively.  
We may include the gravitational potential on the sphere in the Hamiltonian of Eq. (\ref{atomic}).
Using Eq. (\ref{thermog}) for the operator of vertical ($z$-component) velocity of the sphere, ${\cal{O}} = v_z$, we have
\begin{eqnarray}
\label{stokes}
v_{l.t.a.; s.c.} &&= \frac{2}{9} \frac{\rho_{sphere} - \rho_{fluid}}{\eta} gR^2 \nonumber
\\ && = \int dE' ~{\it{p}}_{T}(E') ~ v'_\infty (E') \nonumber
\\ &&=  \int_{0}^{\infty} dT' ~v'_{\infty}(E'(T')) ~{\it{p}}_{T} \big(E'(T')\big) ~ C'_{V}(T') \nonumber
\\ &&+ \int_{{\cal{PT}}E{\cal{I}}} dE' ~ {\it{p_{T}}} (E') ~ v'_\infty (E').
\end{eqnarray}
This equation is exact. Here, $v'_\infty(E')$ is the terminal velocity of the sphere for an equilibrated system of energy $E'$ (for energy densities that lie
outside the ${\cal{PT}}E{\cal{I}}$, the energy $E'$ is equal to the internal energy of the equilibrated system
at temperature $T'$). The first equality in Eq. (\ref{stokes}) is that of Stokes' law. In any given eigenstate of Eq. (\ref{atomic}), the long time average of the sphere velocity $v_z$ (or any other quantity ${\cal{O}}$) 
is equal to its expectation value in that eigenstate.
Eq. (\ref{stokes}) follows from Eq. (\ref{thermog}) and relates the terminal sphere velocity 
(and thus the measured viscosity) in the supercooled liquid to terminal velocities (set by viscosities) in equilibrated liquids at temperatures $T'$. 
Later on, we will derive an approximate form for the viscosity using Eq. (\ref{stokes}) (depicted in Fig. \ref{OTP.}b). 
However, prior to making any assumptions regarding the functional forms of ${\it{p}}_{T}(E')$ and $v'_\infty (E')$ and in line with the general theoretical principle that underlies the current work, we make a general observation. If the energy $E'=E_{melt}$ is the sole singular energy governing the equilibrium behavior (including the velocity $v'_\infty (E')$) then the distribution ${\it{p}}_{T}(E')$ and the viscosity in Eq. (\ref{stokes}) may be a function of the dimensionless ratio $(E-E_{melt})/E$ or, equivalently, of the associated scale free temperature as measured from the melting temperature, $(T-T_{melt})/T$. Thus, an associated universal data collapse of the viscosity 
may occur. As will be elaborated, we have indeed verified that this is the case \cite{nick,nick'} (a collapse occurs with the use of this single scale free temperature). This is clearly seen in Figure \ref{nussinov_collapse} of the current work (reproduced from \cite{nick}). 

\section{Relaxation rates.} 
\label{relaxing}
We next discuss a more general, yet weaker, semi-classical calculation for dynamical quantities in which off-diagonal contributions (in the eigenbasis of $H$) are also omitted at finite times. In such a calculation, the relaxation rates $r_{s.c.}$  from $| \psi \rangle$ to unoccupied states $\{|\phi_m \rangle\}$ generated by a time dependent perturbation $U_{pert.}(t)$ are, similar to Eqs. (\ref{llong}, \ref{thermog}), given by
\begin{eqnarray}
\label{fix}
r_{s.c.}&&= \frac{d}{dt} \sum_{m} |\langle \phi_m| U_{pert.}(t) | \psi \rangle|^{2}= 
 \sum_{n} |c_{n}|^{2} \frac{d}{dt} \sum_{m}  |\langle \phi_m| U_{pert.}(t)| \phi_{n} \rangle|^{2}   \nonumber 
\\ &&\equiv \sum_{n} |c_{n}|^{2} r_{n}  =   \int_{0}^{\infty} dT'   ~r'(T')~{\it{p}}_{T} \big(E'(T')\big) ~ C'_{V}(T') \nonumber
\\ &&+  \int_{{\cal{PT}}E{\cal{I}}} dE' ~ {\it{p_{T}}} (E') ~ \tilde{r}'(E').
\end{eqnarray}
The second equality in Eq. (\ref{fix}) is valid only if off-diagonal terms are neglected \cite{diagonalll}. Here, $r_{n}$ is the relaxation rate of a particular eigenstate $|\phi_{n} \rangle$ to the set of states $\{|\phi_m \rangle\}$ subject to the same time dependent perturbation Hamiltonian 
that defines the evolution operator $U_{pert.}(t)$. With this same perturbation, the temperature dependent relaxation rate $r'(T')$ is that of an {\it equilibrated} solid or liquid at the temperature $T'$ for which the internal energy $E'(T')=E_{n}$. The rate $\tilde{r}'(E') = r_n$ for energies $E'=E_{n}$ in the 
${\cal{PT}}E{\cal{I}}$ (for which $T'=T_{melt}$). Analogous to Eqs. (\ref{lllong},\ref{shorteq}), a relation equivalent to Eq. (\ref{fix}) appears for a calculation with the diagonal density matrix when off-diagonal contributions are omitted; such a calculation leads, once again, to the final equality of Eq. (\ref{fix}). Now here is an important point: {\it the rates $r'(T')$ may be determined from experimental measurements}. Typically, the liquid ($T'>T_{melt}$) equilibrium relaxation is governed by an Eyring type rate \cite{long}, 
\begin{eqnarray}
\label{rrate}
r'(T')= r'(T_{melt}) \frac{T'}{T_{melt}} e^{(\frac{\Delta G(T_{melt})}{k_{B} T_{melt}} - \frac{\Delta G(T')}{k_{B} T'})},
\end{eqnarray} with $\Delta G(T)$ a Gibbs free energy barrier. Because the viscosity is given by $\eta_{s.c} \propto 1/r_{s.c.}$, changes in the hydrodynamic relaxation 
rate with temperature may be evaluated from the corresponding changes in the 
terminal velocity of Eq. (\ref{stokes}). We remark that when $T_{melt} + {\overline{\sigma}}> T' > T_{melt}$ with ${\overline{\sigma}} \ll T_{melt}$, the Gibbs free energy barrier varies weakly, $\Delta G(T') \approx \Delta G(T_{melt})$, and the equilibrium relaxation rate $r'(T') \approx r'(T_{melt})$. 
As may be readily rationalized, e.g., \cite{4He}, the same dominant relaxation times $\tau_{s.c.} = r_{s.c.}^{-1}$ that govern the viscosity are also phenomenologically present in other response functions such as the frequency dependent dielectric response function, $\epsilon(\omega) = \epsilon_{0} (1- i \omega \tau_{s.c.})^{-\overline{\beta}}$ (with an exponent $0<\overline{\beta} <1$)  \cite{DC} associated with a perturbing electric field. Qualitatively, these frequency dependencies emulate a real-time ``stretched exponential'' ($e^{-(t/\tau_{s.c.})^a}$ with $0<a<1$) behavior that may formally be expressed as an integral over a broad distribution $\rho(\tau')$ of relaxations of the $e^{-t/\tau'}$ type. The evaluation of the viscosity that we will shortly embark on does not rely on the neglect of finite time off-diagonal terms as in Eq. (\ref{fix}). This is so as Eq. (\ref{stokes}) enables the determination of the viscosity from long time measurements (when, indeed,
the off-diagonal terms may identically drop due to phase cancellations as in Eqs. (\ref{llong},\ref{thermog})).

\section{The energy as a function of temperature.} 
\label{sec:Energy}

In principle, if we know {\bf (1)} the equilibrium heat capacity $C'_{v}(T')$ and the distribution ${\it{p}}_{T}$ at all temperatures and at all energies in the ${\cal{PT}}E{\cal{I}}$ interval or
{\bf (2)} the distribution ${\it{p}}_{T}$ at all energies then we will be able to directly compute the long time average of all observables ${\cal{O}}$ in the supercooled liquid given their values in the equilibrium system. 

We next review the equilibrium heat capacity $C'_{v}(T')$ in equilibrated solids as it is relevant to the integral in Eq. (\ref{thermog}, \ref{stokes}, \ref{fix}). 
At high $T'$, the heat capacity of a simple equilibrated harmonic solid of $N$ atoms is $C'_v = 3Nk_{B}$ (and $E'(T') = 3 Nk_{B}T'$). As $T'$ is lowered, the heat capacity decreases. In the Debye model (e.g., \cite{AM})
$E'(T') = (3 Nk_{B}T') F_{D}(\frac{T_{D}}{T'})$ with $F_{D}(z)=\frac{3}{z^3} \int_{0}^{z} \frac{t^3~dt}{e^{t}-1}$. At high temperatures, $\lim_{z \to 0}F_D(z) =1$. For many solids, the Debye temperature $T_D$ setting the cross-over scale to the limiting high $T'$ form of the energy $E(T')$ lies in the range $[\frac{T_{melt}}{4} ,\frac{T_{melt}}{2}]$. For many glass formers, $T_{g} \sim 0.7 ~T_{melt} $ \cite{kt} while in metallic glasses, $T_{g} \sim 0.55 T_{melt}$ \cite{metallic-glass}. Thus,  in the equilibrated solids, at temperatures between $T_{g}$ and $T_{melt}$, the heat capacity is {\it nearly constant}. 

The first term in the last equality of Eq. (\ref{thermog}) does not include contributions from the ${\cal{PT}}E{\cal{I}}$ interval. We now turn to an approximate form of the energy in the supercooled liquid that will be of greater utility in calculations that use ${\it{p}}_{T}(E')$ at all energies $E'$ using the second equality of Eq. (\ref{thermog})  (i.e., case {\bf{(2)}} above). The heat capacity of the supercooled liquid must satisfy Eq. (\ref{ECE}). All features of the heat capacity (including its well-known (typical faint) peak near the (calorimetric) glass transition temperature $T_{g}$) must be imprinted in the distribution ${\it{p}}_{T}(E')$. To fully explore these and other properties, the contributions to the energy from both the ${\cal{PT}}E{\cal{I}}$ region and the temperature interval $T'<T_{melt}$ of the equilibrated solid must be included. Usually, the heat capacity of the supercooled liquid does not change markedly in the temperature range $T_{g} < T <T_{melt}$. The internal energy (or enthalpy in systems with fixed pressure) has long been well approximated by a linear
function in $T$ with negligible higher order corrections \cite{stebbins,cvp}. Thus, the difference between (i) the average of the energy $E'$ as computed with $p_{T}(E')$ (i.e., $E$) 
and (ii) the energy of the system at equilibrated system at melting ($E_{melt}$), may be estimated by
\begin{eqnarray}
\label{EC}
(E_{melt}- E(T)) = C (T_{melt}-T).
\end{eqnarray} 
Here, $C$ is an effective average heat capacity of the supercooled liquid. In later sections, we will return to the quantity $C$ and, for simplicity, assume that a ratio 
formed with its aid (Eq. (\ref{wide'})) is nearly constant in the temperature range of empirical relevance. It is important to note that while the ${\cal{PT}}E{\cal{I}}$ might correspond to a narrow or single temperature in the equilibrium system (the melting temperature $T_{melt}$ in an ideal uniform system), it may be associated with a {\it broad range of temperatures} in the supercooled liquid. Thus, the temperature $T_{-}$ at which the energy density of the supercooled liquid is equal to $E_{melt}^{-}/V$ (the lowest energy density in the mixed liquid-solid equilibrium system) may be far below the melting (or liquidus) temperature $T_{melt}$ (at which energy density is $E_{melt}^{+}/V \equiv E_{melt}/V$). Below $T_{-}$, we may anticipate the supercooled system to be dominated by a solid like behavior.

\section{Computing the viscosity via the eigenstate distribution.}
 
We now return to the terminal  velocity relations derived in Section \ref{vlv}. Applying these to physical systems, we note that the terminal velocity of a sphere placed on an equilibrium solid vanishes and similarly the expectation value of the vertical velocity $v_{z}$
of the dropped sphere within low energy solid eigenstates is zero \cite{explain_zero+,latent_vel},
\begin{eqnarray}
\label{vv}
\langle \phi_{n} | v_{z} | \phi_{n}\rangle = 0, ~~ E_{n} < E_{melt}.
\end{eqnarray}
In the terminology of the Introduction, Eq. (\ref{vv}) constitutes an effective {\it ``selection rule''}. 
In the current context, the evolution of the system in response to the applied external shear
(the external gravitational field acting on there sphere)
is trivial for components of the wavefunction that have an energy density lower than that at melting.
In systems with significant latent heat (${\cal{PT}}E{\cal{I}}$) contributions to the terminal velocity, the upper bound on $E_{n}$ of $E_{melt}$ in Eq. (\ref{vv}) should be replaced by the lowest energy in the 
${\cal{PT}}E{\cal{I}}$ region. That is, in such instances, the expectation values $\langle \phi_{n} | v_{z} | \phi_{n}\rangle = 0$
will (in the notation of Section \ref{generalc}) vanish for energies $E_{n}<E_{melt}^{-}$ 
below which the equilibrium system is completely solid and no average long time flow occurs. By contrast to Eq. (\ref{vv}), the expectation value of the velocity, $\langle \phi_{n} | v_{z} | \phi_{n}\rangle$ does not vanish in the liquid-type eigenstates having $E_{n} > E_{melt}$. 
Thus in Eq. (\ref{stokes}) (and similarly in Eq.(\ref{fix})), only
the higher energy ($E_{n}>E_{melt}$) liquid like states $\{|\phi_{n} \rangle\}$ enable hydrodynamic flow that the perturbation may induce
\cite{dosolid}. 
Thus, the integral $\int_{E_{melt}}^{\infty} dE'~{\it{p}}_{T}(E')$ has most of its support in a narrow
region of width $\sigma_T$ near $E_{melt}$ (the high energy tail of ${\it{p}}_{T}(E')$ thins out very rapidly for $E'>E_{melt}$).
Using Eqs. (\ref{stokes},\ref{fix}), we may then estimate the long time velocity or the associated relaxation rate of the supercooled liquid,
$r_{s.c.}(T) \simeq r'(T_{melt}) \int_{E_{melt}}^{\infty} ~{\it{p}}_{T}(E')~ dE'$. 
The viscosity is set by Eq. (\ref{stokes}) or, equivalently, scales with the relaxation time $\tau_{s.c.} =1/r_{s.c.}$. We reiterate that the long time average of the velocity vanishes in the equilibrium solid phase, $v'_\infty(E'(T'<T_{melt}))=0$. (Stated formally, hydrodynamic relaxation due to shear is largely absent in this phase, $r^{hydro}(T'<T_{melt}) =0$). Since the long time average of the velocity within the equilibrated system vanishes at low energies 
($v'_\infty(E'(T'<T_{melt}))=0$), we see from Eq. (\ref{stokes}) that the viscosity $\eta_{s.c.}$ of the supercooled liquid must satisfy
\begin{eqnarray}
\label{vis}
\eta_{s.c.}(T) \lesssim
\frac{\eta'(T^{+}_{melt})}{  \int_{E_{melt}}^{\infty} ~{\it{p}}_{T}(E')~ dE'} \equiv \overline{\eta}(T).
\end{eqnarray}
Here, $\eta'(T^{+}_{melt})$ is the viscosity of an equilibrated liquid at temperatures infinitesimally above melting. 
Eq. (\ref{vis}) will become an equality if the ${\cal{PT}}E{\cal{I}}$ contributions to $v_{l.t.a.; s.c.} $
(associated with a long time velocity $v'_\infty(E')$ within the equilibrium spatially coexisting liquid and solid regions at the melting temperature) are ignored in Eq. (\ref{stokes}). This is indeed what we will largely assume. Unless stated otherwise, in what will follow, we will set $ \eta_{s.c.}(T) = \overline{\eta}(T)$. Due to the constraint of Eq. (\ref{constraint}), the energy distribution ${\it{p}}_{T}(E')$ will have an average equal to (and, typically, be centered about) the system energy $E$. By Eq. (\ref{constraint}), when $T$ is far smaller than $T_{melt}$ (and the corresponding energy density far lower than that of the equilibrium system at melting), the denominator in Eq. (\ref{vis}) will become a near vanishing integral, {\it leading to a very large viscosity}. While the viscosity becomes large at these low temperatures $T$ , the attendant thermodynamic and other static observables of Eq. (\ref{thermog}) need not exhibit a striking change. 

We conclude this section by underscoring the physical aspects of our results and their viable relations to earlier approaches. As intuitively suggested, rare events may contribute significantly to motion in glasses. In one form or another, many classical approaches to the glass transition, e.g., \cite{ag,ag1,ag2,ag3,CG,ejcg,elm,jam,jam',gt1,sri,gt2,gt3,gt4,gt5,gt6,moore,myega,wg,mdd,avoided1,avoided2,avoided3} visualize such events (rare motions in jammed systems \cite{jam,jam'} or unlikely transitions within an energy landscape, etc.). In our framework, these scarce occurrences that enable flow {\it need not} directly correspond to final supercooled liquid states of high energies $\langle \psi | H | \psi \rangle$. Rather, as is seen from 
the {\it{``selection rule''}} of Eq. (\ref{vv}), in order to exhibit appreciable motion, in the eigenstate decomposition of $| \psi\rangle$ (Eq. (\ref{dec})), the cumulative weight of the states of energies $E_{n} > E_{melt}$ must be high. To properly account for motion in low temperature supercooled liquids, we need to know how to ``count'' and sum contributions from the rare states that enable dynamics. 
In effect, Eqs. (\ref{stokes}, \ref{vv}) demonstrate that the distribution ${\it{p}}_{T}(E')$ weighs these states in precisely the correct way.

\section{The probability distribution.} 
\label{sec:prob}

We now turn to the probability density that appears in all of our equations.
In this section, we will suggest that the probability distribution ${\it{p}}_{T}(E')$ is a Gaussian (Eq. (\ref{Gauss''}))
of a finite width (Eq. (\ref{wide'})). 
The probability density ${\it{p}}(E')$ has two curious physical characteristics triggered by the supercooling: 
\newline
{\bf{(i)}} ${\it{p}}_{T}$ {\it cannot be a delta function} as function of the energy density $E'/V$. This assertion is readily established via a simple proof by contradiction. To this end, suppose that the
opposite is true and that system is spectrally confined to a narrow energy interval $\Delta E$ (i.e., only a narrow range of energies appear in the spectral decomposition of Eq. (\ref{dec})). In such a situation, the long time averages of general observables ${\cal{O}}$ as computed by the righthand side of 
Eq. (\ref{shorteq}) will equal to the microcanonical average of Eq. (\ref{mc}) with the aforementioned small $\Delta E$. However, if that is the case, then it follows (by definition) that general expectation values
 $\langle {\cal{O}} \rangle$ must be equal to their counterparts when computed within the microcanonical ensemble (i.e., to their values as appearing on the lefthand side of Eq. (\ref{etheq})) \cite{explain_mc_}. If ${\it{p}}_{T}$ is a delta function distribution in the energy density then, inevitably, the system will be in equilibrium insofar as any probe measuring ${\cal{O}}$ can attest. The supercooled liquid is, however, radically different from an equilibrated liquid or equilibrated solid. Thus, we see that the lack of thermalization of the supercooled liquid implies a {\it spread of energy densities} in the spectral decomposition of $| \psi \rangle$.  \newline
{\bf{(ii)}} By its non-equilibrium nature and as it is generated by external supercooling, it may be impossible to derive ${\it{p}}_{T}(E')$ by thermodynamic considerations associated with the Hamiltonian $H$ alone. 

In the subsections that follow, we will first recall the Gaussian probability distribution in equilibrated systems (\ref{ref:equil}). We will then turn to our general non-equilibrium systems in Section \ref{ref:un}.  We will explain why a Gaussian distribution is anticipated from simple Shannon entropy considerations (\ref{EMsec}) also for our non-equilibrium system; the Gaussian form that we will arrive at exhibits an extensive standard deviation of the distribution ${\it{p}}_{T}(E')$ (as mandated by property {\bf{(i)}} above). We will explain why such a finite standard deviation enables a sharp value of the internal energy (\ref{sharp}). Next, we will describe how an effective non-local Hamiltonian may be generated by supercooling; this non-locality renders void the usual arguments concerning ${\cal{O}}(N^{1/2})$ fluctuations of the energy and other extensive quantities (\ref{sec:non-locaiity}). To concretely elucidate these aspects, we will examine a toy model in which extensive fluctuations appear (\ref{e-fluc}). We conclude Section \ref{ref:un} by underscoring qualitative experimental consequences that are associated with an extensive standard deviation (\ref{nontrivfinal}). These features are in agreement with experimental observations. 

\subsection{Equilibrated systems.}
\label{ref:equil}
To motivate the simplest possible functional form of ${\it{p}}_{T}(E')$ consistent with features {\bf{(i)}} and {\bf{(ii)}} and to better appreciate the quintessential character of this distribution for general supercooled fluids, we first briefly review the energy distribution in equilibrated systems. We consider the standard case of equilibration between a liquid and a heat bath ($B$) such that the combined (liquid-bath) system
has an energy $E_{tot}$. The equilibrium probability density ${\it{p}}^{eq}_{T} (E') = P(E_{B} = E_{tot} - E')  \propto e^{S(E_{B} = E_{tot} - E')/k_{B}}$ with $S=S_{B} +S_{liquid}$, the entropy of the combined liquid and thermal bath system. Taylor expansion of $S$ in the energy of the small liquid system $E' \ll E_{tot}$ yields that $P(E_{B} = E_{tot} - E')=\tilde{{\cal{N}}} e^{-\frac{1}{2  k_{B}T^2 C_{v}} (E' -E(T))^{2}}$ where ${\tilde{\cal{N}}}$ is a normalization constant. Here, $E'$ is the energy of the liquid and $E(T)$ is the internal energy of the liquid at temperature $T$. A term linear in $E'$ is absent from the Taylor expansion of the entropy. This is consistent with the fact that the temperatures of the equilibrated bath and the liquid are equal, $\partial S_{B}/\partial E_{B} = \partial S_{liquid}/\partial E'(= 1/T)$. The function $P(E_{B} = E_{tot} - E')$ is the probability for the liquid to have an energy $E' $ or, equivalently, for the bath to have an energy $(E_{tot} - E')$-  {\it it is not} the probability that a particular state be of energy $E'$ (the latter is, of course, proportional to a Boltzmann weight $e^{-E'/(k_{B} T)}$). The sum of Botlzmann weights over all such states is ${\it{p}}^{eq}_{T}(E')=\sum_{m} e^{-(E_{m} -F)/(k_{B} T)} \delta(E'-E_{m})$, where $F= - k_{B} T \ln Z$ is the free energy. Thus, as is well known, the energy probability distribution is a Gaussian,
\begin{eqnarray}
\label{Gauss'}
{\it{p}}^{eq}_{T}(E') = \frac{1}{\sqrt{2 \pi (\sigma^{eq}_T)^2}} e^{-\frac{(E'-E'(T))^{2}}{2 (\sigma^{eq}_T)^2}}.
\end{eqnarray}
The Gaussian form of Eq. (\ref{Gauss'}) also follows from maximization of the entropy at $E'=E(T)$. We will shortly return to maximization arguments. In thermalized systems, 
\begin{eqnarray}
\label{seq}
(\sigma^{eq}_T)^2 = k_{B} T^2 C'_v.
\end{eqnarray}
Since the internal energy $E'$ and thus the heat capacity $C'_v$ are extensive (i.e., scale with the system size $N$), in equilibrium systems, the dimensionless ratio 
\begin{eqnarray}
\label{eq_ratio}
A_{equilibrium} &&\equiv \Big( \frac{\sigma^{eq}_{T}}{C'_{v} T} \Big)_{equilibrium} \nonumber
\\ &&=    \sqrt{\frac{k_B}{C'_v}} \rightarrow_{V \to \infty} 0
\end{eqnarray}
tends to zero in the thermodynamic limit. The spread of energies $\sigma_T$ scales subextensively with the system size and the probability distribution for the energy density is a delta function about $E'(T)/V$. This delta function form in the thermodynamic limit is consistent with our arguments thus far: thermalization corresponds to a narrow distribution of energy densities. When the Hamiltonian is a sum of decoupled terms (and the Boltzmann probability distribution accordingly becomes a product of independent (not necessarily identical) probabilities), Eq. (\ref{Gauss'}) reduces to the central limit theorem of statistics of large but finite size systems. In its most common formulation, the central limit theorem \cite{CLT} states that the arithmetic average $\overline{X}$ of $\tilde{n}$ {\it independent} variables $\overline{X} = \frac{1}{\tilde{n}} \sum_{i=1}^{\tilde{n}} X_{i}$ follows a Gaussian distribution with a relative error $\frac{\sigma_{\overline{X}}}{\overline{X}} \sim {\cal{O}}({\tilde{n}}^{-1/2})$ where $\sigma_{\overline{X}}$ is the standard deviation of $\overline{X}$. For an equilibrium system of size $N$, the pertinent number of terms $n$ in the sum for the energy (and general quantities $X$) scales with the system size $N$ and relative ${\cal{O}}(N^{-1/2})$ fluctuations result for $A_{equilibrium}$ of Eq. (\ref{eq_ratio}). 

\subsection{Unequilibrated supercooled liquids.}
\label{ref:un}
We now examine our case of the supercooled fluid. As throughout this work, we will aim to invoke the smallest number of assumptions. Instead, we will appeal to general principles and draw analogies (and differences) with the derivation of the probability distribution in equilibrium statistical mechanics (that we reviewed in subsection \ref{ref:equil}).

\subsubsection{Width of the probability distribution for achieving different eigenstates.}
To {\it thwart equilibration} (see characteristic {\bf{(i)}}), in supercooled liquids, the dimensionless width of the probability density width {\it must} be non-zero, namely,
\begin{eqnarray}
\label{wide'}
\overline{A} \equiv \Big( \frac{\sigma_T }{ CT} \Big)_{s.c.} = \frac{\sigma_T (E_{melt} - E(T))}{T(T_{melt}-T)}>0.
\end{eqnarray} 
Here, we invoked Eq. (\ref{EC}) for the definition (and scale) of $C$. 
In the simplest approximation, for temperatures $T_{g} < T<T_{melt}$ where viscosity data is available, 
the ratio of Eq. (\ref{wide'}) is $T$ independent just as it is in finite size equilibrated systems if $C_v$ is nearly constant. This dimensionless {\it non-equilibration parameter} ${\overline{A}}$ depends on the cooling protocols. Non-zero values imply deviations from equilibrium; as such, this parameter constitutes a dimensionless
measure of the effect of supercooling (i.e., the widening of ${\it{p}}_{T}$). 
When fitting experimental data with the predictions of the current work \cite{me1}, it was found that in all known types of supercooled liquids (silicates, metallic, and organic liquids) \cite{nick,nick'}, the small dimensionless fraction $\overline{A}$ did not vary much.The nonzero value of $\overline{A}$ superficially emulates the spread of the fluctuations $A_{equilibrium}$ in an equilibrated system of a {\it finite} volume $V$. 
The extended distribution of the energy densities (the non-vanishing dimensionless constant 
${\overline{A}}$ in Eq. (\ref{wide'}))
is intimately tied to the deviation of structure of the supercooled liquid from that of equilibrated solids and crystals. Of course, at high enough $T$ close to the initial temperature of the system prior to supercooling and/or in those instances where the thermalization rate is high relative to the cooling rate, the width of ${\it{p}}_{T}(E')$ should be small. 

\subsubsection{Entropy Maximization with the standard deviation as a parameter leads to a Gaussian eigenstate distribution of finite width.} 
\label{EMsec}
The precise form of the distribution ${\it{p}}_{T}(E')$ depends on system details that go even beyond the unperturbed system Hamiltonian $H$(characteristic {\bf{(ii)}}). In the absence of specifics, one may attempt to employ ``Entropy Maximization'' in order to obtain ``the least biased guess'' concerning ${\it{p}}_{T}(E')$ \cite{jaynes}. In our case, the entropy is that derived from the distribution ${\it{p}}_{T}(E')$. As is well known, the distribution that maximizes the Shannon entropy, 
\begin{eqnarray}
\label{ME}
{\sf{H}}_{S} = - \int dE' {\it{p}}_{T}(E') \ln {\it{p}}_{T}(E'),
\end{eqnarray}
with the constraint that the distribution ${\it{p}}_{T}(E')$ has a fixed standard deviation $\sigma_T$ is a Gaussian. This suggests that in the non-equilibrated system, the probability distribution will indeed be of a form similar to that of Eq. (\ref{Gauss'}), centered about the energy of the supercooled liquid $E(T)$, 
 \begin{eqnarray}
 \label{Gauss''}
 {\it{p}}_{T}(E') = \frac{1}{\sqrt{2 \pi \sigma_T^2}} e^{-\frac{(E'-E(T))^{2}}{2 \sigma_T^2}}.
 \end{eqnarray}
 That is, if we invoke the non-vanishing value of $\sigma_T$, the most general probability distribution that maximizes the Shannon entropy of Eq. (\ref{ME}) 
 is given by Eq. (\ref{Gauss''}) \cite{Gauss-remind}. Unlike the equilibrium distribution of Eq. (\ref{Gauss'}), the probability density of the supercooled liquid ${\it{p}}_{T}(E')$ will have a larger width $\sigma_T$ satisfying Eq. (\ref{wide'}) \cite{minor}. Our anticipation for a Gaussian appearing in various systems that do not exhibit special constraints (i.e., those for which the information theoretic ``Entropy Maximization'' principle may be applied) is in accord with the currently limited numerical data on quantum systems. Indeed, earlier work examined (an analog of) the distribution ${\it{p}}_T(E')$ for a quenched one-dimensional hard-core bosonic system \cite{polkovnikov1,polkovnikov2}. Away from rare exactly solvable points, the distribution was, indeed, numerically well approximated by a Gaussian. To close the circle of our ideas thus far and make contact with the discussion at the beginning of the current section, we briefly review the canonical equilibrium case. If the standard deviation $\sigma_T$ is constrained to its fixed equilibrium value then the maximization of the entropy of Eq. (\ref{ME}) will lead to a Gaussian of the vanishing relative energy width in 
Eq. (\ref{eq_ratio}) as it consistently does in these equilibrated systems \cite{remind}. Upon supercooling, locally stable low energy microstates (so-called ``inherent structures'' \cite{inherent}) typically form. If this set of states is very special then our general Entropy Maximization principle may falter. 
Similarly, if certain configurations persist to lower temperatures or appear at specific temperatures then these may further amend ${\it{p}}_{T}$.

\subsubsection{Sharpness of the measured energy in realizable physical states.}
\label{sharp}
As we emphasized earlier, we do not have a single pure state and, instead, need the full density matrix. By diagonalizing the density matrix $\rho$, we find the probabilities $p_{h}$ for obtaining the different pure states  $\{ | \psi_{h} \rangle \}$. As we emphasized in Section \ref{lta-sec}, the pure states $\{ | \psi_{h} \rangle \}$ generally differ from the eigenstates of the Hamiltonian $H$. By its definition of Eq. (\ref{pt+}), the distribution ${\it{p}}_{T}(E')$ is the probability density for obtaining {\it eigenstates} of energy $E'$. It is the weight of obtaining an energy $E'$ in a {\it spectral decomposition} of  the supercooled state (Eq. (\ref{dec})). The spread of the distribution $ {\it{p_{T}}} (E') \to \rho_{n'n'}$ (see Eqs. (\ref{lllong}, \ref{shorteq})) {\it does not} imply a corresponding spread of energies $\langle \psi_{h} | H | \psi_{h} \rangle$ in the physically pertinent states $\{| \psi_{h} \rangle\}$ (i.e., those of sufficiently high probability). To underscore this issue, we consider 
a trivial example- that of a single state. Given any particular state $| \psi \rangle$, there is a non-trivial distribution ${\it{p}}_{T}(E')$ given by Eq. (\ref{pt+}). However, for any such single $|\psi \rangle$, there is a unique value of the energy $E=\langle \psi | H | \psi \rangle$ (i.e., the probability of having the energy equal to $E$ is trivially unity). 

In what follows, we explicitly consider what occurs after supercooling ceases and the system remains in contact with the environment (``constant temperature reservoir'') for a long time so that its temperature/energy density is well defined. In such a case, the energy fluctuations in the pertinent states $| \psi_{h} \rangle$ will be small and we will indeed have that $E_{h}= \langle \psi_{h} | H | \psi_{h} \rangle$ will not vary significantly between
states $| \psi_{h} \rangle$ of high probability (for otherwise, the notion of a well defined temperature is ill defined). As in the standard equilibrium thermodynamics, setting $\beta$ (``an inverse temperature'') to be a Lagrange multiplier forcing the computed energy to be equal to its measured value (the inverse temperature of equilibrium thermodynamics), the standard thermodynamic relations are obtained. In particular, the variance of the energy fluctuations may, similar to Eq. (\ref{seq}), be given by $\sigma_E^{2} = C_v/(k_{B}  \beta^{2})$. We now briefly review known facts and further explain how our approach leads to these results. If we denote by $p_h$ the probability of an energy $E_{h}=\langle \psi_{h} | H | \psi_{h} \rangle$ then if the system is to have a fixed energy then (similar to the logic that we applied above) maximizing the entropy $(-  \sum_{h} p_h \ln_{h} p_h)$ subject to the condition of fixed average energy $\sum_{h} p_{h} E_{h}$ provided by the total energy exchange with an external heat bath will, as is well known, lead to the Boltzmann probability distribution $p_{h} = e^{-\beta E_{h}}/\sum_{h'} e^{-\beta E_{h'}}$. The sum of the probabilities $p_h$ over all states sharing the same energy becomes the equilibrium distribution for a measurement of the energy in all states that may be physically realized. That is, Eqs. (\ref{Gauss'}, \ref{seq}, \ref{eq_ratio}) describe the probability of obtaining energies $E$ in the physical states $| \psi_{h} \rangle$. The (Boltmzann) probability distribution $p_{h}$ and its sum $\sum_{h} p_{h} \delta(E-E_{h})$ (emulated, in the thermodynamic limit, by the sharp Gaussian of Eqs. (\ref{Gauss'}, \ref{seq}, \ref{eq_ratio})) should not be confused with the probabilities ${\it{p}}_{T}(E')$ associated with the decomposition of the states $| \psi \rangle$ into the eigenstates of $H$. Thus, we see the similarity in deriving the Boltzmann type probabilities for the energies in the physically accessible pure states (as in the standard relations of Eqs. (\ref{Gauss'}, \ref{seq}, \ref{eq_ratio})) and the broad Gaussian probabilities associated with the eigenstate decomposition of such states (Eqs. (\ref{wide'}, \ref{Gauss''})). Both of these, very different, Gaussian probability distributions arise from the Entropy Maximization principle. In \cite{nick2}, we will provide numerical results and analysis demonstrating the sharpness of the energy and its Gaussian character. It is important to repeat and emphasize that although $p_{h}$ is of a Boltzmann form,
the density matrix is {\it not} that of the canonical ensemble of equilibrium thermodynamics ($\rho \neq \rho_{eq}= \exp(-\beta H)/(Tr(\exp(-\beta H)))$). This is so as $|\{\psi_{h} \rangle\}$ are {\it not} eigenstates of the Hamiltonian $H$.

\subsubsection{Non-adiabatic cooling leads to a non-local Hamiltonian.}
\label{sec:non-locaiity}

For a {\it non-adiabatic} $\tilde{H}(t)$ (see Section \ref{ScS}) that embodies the supercooling process, the Hamiltonian 
\begin{eqnarray}
\label{ht*}
{\cal{H}} \equiv \tilde{U}^{\dagger}(t_{initial}, t_{final})  H \tilde{U} (t_{initial}, t_{final})  \nonumber
\\ =   e^{i (\Delta t) {\tilde{H}} (t_{initial}) /\hbar} \cdots e^{i   (\Delta t) \tilde{H}(t_{final} - 2 \Delta t )/\hbar} e^{i  (\Delta t) \tilde{H}(t_{final}  - \Delta t) /\hbar} \nonumber
\\ \times H  e^{i (\Delta t) \tilde{H}(t_{final}  - \Delta t) /\hbar} e^{i  (\Delta t)  \tilde{H}(t_{final} - 2 \Delta t )/\hbar}   \cdots  e^{-i  (\Delta t) {\tilde{H}} (t_{initial})/\hbar}
\end{eqnarray}
describes, in the Heisenberg picture, the supercooled system at times $t> t_{final}$. In the initial equilibrium state $| \psi (t_{initial}) \rangle$ the Hamiltonian $H$ has a small variance. As ${\cal{H}}$ does not commute with $H$, the relative size the standard deviation of ${\cal{H}}$ when computed in the state $| \psi(t_{initial}) \rangle$ may be non vanishing (Eq. (\ref{wide'})). That is, the state $| \psi(t_{initial}) \rangle$ is associated with the density matrix constructed from $H$. However, as we stress, this state is no longer a (near) eigenstate of ${\cal{H}}$; consequently the variance of ${\cal{H}}$ in $| \psi (t_{initial}) \rangle$ need not be vanishingly small. 

In standard systems with local interactions, the variance of the energy scales as ${\cal{O}}(N^{1/2})$In what follows, we broadly wish to highlight that,
in our case, ${\cal{H}}$ will {\it not be a local Hamiltonian} even if the initial general $H$ is local. 
To illustrate this property, we may recursively invoke the Baker-Campbell-Hausdorff formula, 
\begin{eqnarray}
\label{ht**}
e^{A} B e^{-A} = B + [A,B] + \frac{1}{2!} [A,[A,B]]  \nonumber
\\ + \frac{1}{3!} [A,[A,[A,B]]] + \cdots,
\end{eqnarray}
in Eq. (\ref{ht*}) at different times $t = t_{initial}, t_{initial} + \Delta t, \cdots, t_{final} - 2 \Delta t, t_{final}- \Delta t $ (i.e., setting $B=H$ and $A= (\Delta t)  \tilde{H}(t) /\hbar$ at consecutive times $t$). To obtain formally explicit forms, we may express the many body Hamiltonian of Eq. (\ref{atomic}) and the cooling Hamiltonian ${\tilde{H}}(t)$ in second quantized forms. 
Regardless of the specific initial $H$ (whether that of Eq. (\ref{atomic}) of that of other more theoretical models), we observe that the resulting Hamiltonian ${\cal{H}}$ {\it need not capture the spatial locality} of the coordinates in $H$. The commutators $[{\tilde{H}}(t), H] \neq 0$ (and $[\tilde{H}(t), {\tilde{H}}(t')] \neq 0$) do not vanish at general times $t_{initial} \le t \le t_{final}$ (and at $t' \neq t$, respectively). As we emphasized earlier, this non commutativity is required in order to change the energy and lower the system temperature. From Eqs. (\ref{ht*}, \ref{ht**}) we see that this property further ensures that even if both $H$ and $\tilde{H}$ are sums of spatially local (or nearly local) terms (that drop off rapidly with distance),
long range interactions will generally appear in the final Hamiltonian ${\cal{H}}$. Thus, as we remarked above, arguments concerning ${\cal{O}}(N^{1/2})$ fluctuations cannot be blindly adopted from the standard case of equilibrated systems with local interactions \cite{brief}. 

\subsubsection{Energy fluctuations in a harmonic oscillator dual.}
\label{e-fluc}

By dimensional analysis, we anticipate that when the standard deviation $\sigma_T$ of $H$ is evaluated in a general supercooled state $| \psi \rangle$
(which is not a (near) eigenstate of $H$), it will scale with the energy (or temperature) as $\sigma_T = {\cal{O}}(E)(= {\cal{O}}(CT))$. That is, we expect that the ratio $\overline{A}$ in Eq. (\ref{wide'}) will, approximately, be nearly constant over a range of energies. We now briefly provide a toy example of a single particle harmonic oscillator in one dimension in which this is indeed realized. We briefly motivate the use of such a system
as a proxy to the far more complex real many body system governed by Eq. (\ref{atomic}) for a narrow range of energies (much unlike a simple harmonic oscillator, the energy levels in the many body system typically become denser as energy increases). Nevertheless, if in a narrow energy interval of interest at a given temperature $T$, the spectra of the many body systems set by both the original Hamiltonian $H$ and its evolved counterpart ${\cal{H}}$ formed by supercooling exhibit a near constant separation between consecutive energy levels, then we may describe these systems by effective isomorphic  ``dual'' Hamiltonians that are those of a single particle one-dimensional harmonic oscillators, $H_{dual}= \frac{p^{2}}{2m} + \frac{1}{2} m \omega^{2} x^{2}$ and ${\cal{H}}_{dual}= \frac{p^{2}}{2M}  + \frac{1}{2} M \Omega^{2} x^{2}$. Thus, we consider the case where, in a certain interval of energies, all of the matrix elements in the many body $\{| \phi_n \rangle\}$ basis (for both the $H$ and ${\cal{H}}$ Hamiltonians) and those of the dual single particle Hamiltonians $H_{dual}$ and ${\cal{H}}_{dual}$  (in the eigenbasis $\{ | \phi^{dual}_{\overline{n}} \rangle \}$ of the harmonic oscillator Hamiltonian $H_{dual}$) are in a one-to-one correspondence.
Within the eigenstates $\{ | \phi^{dual}_{\overline{n}} \rangle \}$  of $H_{dual}$ (with $\overline{n}=0,1,2, \cdots$) the first moments of ${\cal{H}}_{dual}$ are, trivially \cite{full-off-diag},
\begin{eqnarray}
\label{moments+}
\langle \phi^{dual}_{\overline{n}} | {\cal{H}}_{dual}| \phi^{dual}_{\overline{n}} \rangle && = \frac{1}{4} (2 \overline{n}+1) \hbar \omega \Big( \frac{m}{M} + \Big( \frac{\Omega}{\omega} \Big)^{2} \frac{M}{m} \Big), \nonumber
\\ \langle \phi^{dual}_{\overline{n}} |  {\cal{H}}_{dual}^{2} | \phi^{dual}_{\overline{n}} \rangle &&= \frac{3}{16}  
(2 \overline{n}^{2} + 2 \overline{n} +1)  \nonumber
\\&& ~ \times \Big(\hbar \omega \Big)^{2}  \Big( \Big(\frac{m}{M} \Big)^{2} + \Big(\frac{M}{m} \Big)^{2} \Big(\frac{\Omega}{\omega} \Big)^{4} \Big)\nonumber 
\\ && ~ + \frac{1}{8} (2 \overline{n}^{2} + 2 \overline{n} -1)  \Big(\hbar \Omega \Big)^{2}, \nonumber
\\  && \cdots . 
\end{eqnarray}
The thermodynamic limit of the original many body system describing $H$ (and the associated ${\cal{H}}$) corresponds to very high energy levels $\overline{n} = {\cal{O}}(\frac{E}{\hbar \Omega}) \ggg 1$ in the representation of the one particle systems. In this limit, the variance $(\langle \phi^{dual}_{\overline{n}} |  {\cal{H}}_{dual}^{2} | \phi^{dual}_{\overline{n}} \rangle - \langle \phi^{dual}_{\overline{n}} | H| \phi^{dual}_{\overline{n}} \rangle^{2})$ grows as ${\cal{O}}(\overline{n} ^{2})$. Thus, the standard deviation of ${\cal{H}}_{dual}$ as measured in these states scales as $\overline{n}$. As the energies and matrix elements of the dual system are in a one to one correspondence to those of the many body system, the variance of ${\cal{H}}_{dual}$ computed in the states $| \phi^{dual}_{\overline{n}} \rangle$
is equal to that of ${\cal{H}}$ computed in the many body states $| \phi_{n} \rangle$. Thus, the standard deviation of  the many body ${\it{p}}_{T}$ scales with the energy $E$ or, equivalently, with the product $CT$. (For the cartoon harmonic oscillator Hamiltonians, at large $\overline{n}$, by the classical equipartition theorem, the energy is exactly linear in $T$.) Thus, in accord with Eq. (\ref{wide'}), the standard deviation $\sigma_T$ associated with the state $| \psi \rangle$ will, generally, scale as the energy $E$.
To repeat our discussion above, we may consider highly excited states of a single harmonic oscillator (or other potential) to qualitatively emulate the spacing of levels in a many body system. The single coordinates in the above cartoon $H_{dual}$ and ${\cal{H}}_{dual}$ ``do not know'' of
the spatial coordinates in the many body system yet if the expectation values and matrix elements  
in this dual one-dimensional (simple harmonic oscillator or other) system are the same as those in the original many body system then the final results 
for distribution ${\it{p}}_{T}(E')$ are identical. If, in the eigenbasis of $H_{dual}$, the off-diagonal matrix elements of ${\cal{H}}_{dual}$ are small (leading to a small finite $\overline{A}$ in Eq. (\ref{wide'})) then higher order cumulants of ${\it{p}}_{T}$ will become progressively smaller. Beyond the above contrived harmonic oscillator or other trivial systems that we can solve for, no exact equalities may be derived. The above harmonic oscillator example was only introduced to illustrate that, as a matter of principle, Eq. (\ref{wide'}) may be realized for non-adiabatic systems. 

\subsubsection{Non-trivial consequences of the broad distribution ${\it p}_{T}$: lack of long range spatial structure and non-uniform dynamics.}
\label{nontrivfinal}

Formally, the non-vanishing standard deviation $\sigma_T$ is the ``uncertainty'' associated with $H$ in the state $| \psi \rangle$ (or an ensemble of such states in the probability density matrix formalism) when computed with the distribution ${\it{p}}_{T}$. As we discussed in Section \ref{lta-sec}, this ``uncertainty'' is trivially set by the long time average of the squared fluctuations. This uncertainty has important physical implications for the system dynamics (including the viscosity) and structure. A significant $\sigma_T$ cannot appear in a pristine state in which all electronic and ionic position and momenta have negligible uncertainty. This is so as if little uncertainty exists in the particle locations and momenta about their average at long times then 
the Hamiltonian of Eq. (\ref{atomic}) will (by comparison to its expectation value) exhibit a negligible uncertainty $\sigma_T$. Thus, it follows that, if $\sigma_T$ is not negligible then we {\it must} have positional and/or momenta {\it uncertainties}. Indeed, a positional uncertainty implies a departure from global ordered structures (in agreement with more detailed arguments to be presented in Appendix \ref{space.}). Therefore, the amorphous structure of glasses may be regarded as a consequence of the non-adiabatic character of the supercooling that triggers a broadening of the energy densities present in the eigenstates comprising the system state (Eq. (\ref{dec})). Similarly, a broad distribution of momenta requires that the particle motion is not spatially uniform (see Section \ref{sec:dh}). This uncertainty cannot be generated by random local thermal fluctuations about an ordered structure such as a crystal. This is so as in a system with local interactions, such local random fluctuations will not yield an extensive uncertainty $\sigma_{T}$ \cite{brief}. As we explained in Section \ref{sec:non-locaiity}, supercooling leads to an effective non-locality. Experimentally, heterogeneous energy distributions have been observed by various probes including dynamic force microscopy (see, e.g., \cite{yhliu} in which an {\it apparent relative energy variation} equal to $0.12$ was seen in a metallic glass former). At best, these experimental measurements may only be seen as a qualitative proxy to Eq. (\ref{wide'}). Further specific details concerning $E(T)$ and ${\it{p}}_{T}(E')$  beyond the general notions outlined here will be provided elsewhere \cite{nick2}. 

We conclude this section by reiterating the central qualitative differences between standard equilibrium statistical mechanics and the non-equilibrated supercooled liquid. Because the density matrix $\rho$ greatly differs from that associated with a stationary eigenstate and is composed of eigenstates having a wide range of energies, during a time interval of size $\Delta t$, it evolves non-trivially under the time evolution operator $e^{-i H \Delta t/\hbar}$ (i.e.,in the decomposition of Eq. (\ref{dec}), eigenstates of different energies evolve differently when acted on by this operator). This evolution, once again, reflects the fact that the system is out of equilibrium and thus its ``state'' (both figuratively and literally) may keep changing non-trivially. That is, as we repeatedly underscored, the supercooled liquid is not in a canonical textbook type ``quasi-stationary equilibrium state''. By examining how many states share the same distribution of Eq. (\ref{pt+}), it is seen that by contrast to the equilibrium case of eigenstates constrained to an infinitesimally narrow energy width, there is a far greater ``{\it extensive configurational entropy}'' associated with
distributions ${\it{p}}_{T}$ having a finite $\overline{A}$ \cite{ccc} that span a far larger range of energy densities.

\section{The viscosity and relaxation times below melting.}
\label{sec:central} From Eqs. (\ref{wide'}, \ref{Gauss''}), we find the following expression for the viscosity,
\begin{eqnarray}
%\begin{equation}
\label{geoni}
%\boxed{
%\begin{aligned}
\boxed{\eta_{s.c.}(T) = \frac{\eta_{s.c.}(T_{melt})}{erfc \Big( \frac{E_{melt}-\langle E \rangle}{\sigma_{T} \sqrt{2}} \Big)}  = \frac{\eta_{s.c.}(T_{melt})}{erfc \Big( \frac{T_{melt}-T}{\overline{A} T
\sqrt{2} } \Big)}.}
%\end{aligned}}
  %\nonumber
%\\ \hspace*{-10cm} 
%\frac{\eta(T_{melt})~ \overline{\sigma} ~e^{\frac{(T_{melt}-T)^{2(1+ \delta)}}{(2 \overline{\sigma}^{2})}}}
%{{\sqrt{2 \pi}(T_{melt}-T)}},
%\end{equation}
\end{eqnarray}
The first equality of Eq. (\ref{geoni}) follows when Eq. (\ref{Gauss''}) is plugged into Eq. (\ref{vis}) (and Eq. (\ref{vis}) appears as an equality instead of a more general bound when the ${\cal{PT}}E{\cal{I}}$ contributions in Eq. (\ref{stokes}) are non-zero); if long time flow were to hypothetically occur for solid-type states this bound would be further strengthened. To obtain  the final result of Eq. (\ref{geoni}), we inserted Eqs. (\ref{EC},\ref{wide'}). We remark that our derivation of Eq. (\ref{geoni}) holds so long as the ratio $\overline{A}$ of Eq. (\ref{wide'}) is $T$ independent (i.e., even if the average heat capacity $C$ satisfying Eq. (\ref{EC}) is temperature dependent). In the approximation of Eq. (\ref{vis}), valid for a Gaussian ${\it{p}}_{T}$
when ${\overline{\sigma}}_T  \equiv \sigma_T/C \ll T_{melt}$, at $T=T_{melt}$, a ``half'' of the distribution ${\it{p}}_{T}$ includes localized solid states while the other half is comprised of delocalized liquid states (see Fig. \ref{eigen.}b), the relaxation rate of the equilibrated liquid is double that of the supercooled fluid. Accordingly, the viscosity of the supercooled liquid at the melting (or, more precisely, liquidus) temperature $T=T_{melt}$ is 
double that of the equilibrated liquid just above melting, i.e., $\eta_{s.c.}(T_{melt}) \approx 2 \eta'(T_{melt}^{+})$. 
When the ${\cal{PT}}E{\cal{I}}$ contributions in Eq. (\ref{stokes}) are included, the more detailed inequality of Eq. (\ref{vis}) 
leads to another prediction of the theory,
\begin{eqnarray}
\label{boundratio}
\frac{\eta_{s.c.}(T_{melt})}{\eta'(T_{melt}^{+})} \lesssim 2.
\end{eqnarray}
The prediction of Eq. (\ref{boundratio}) may be tested by noting that in the equilibrium system the viscosity follows an Arrhenius form (as we will discuss
in Section \ref{vmt}), i.e., $\eta'(T') \simeq \eta_{0} \exp(\Delta G/(k_{B} T'))$ with $\Delta G$ an approximately temperature independent
free energy barrier. Thus, one may examine a fit of the viscosity of the glass forming liquid at high temperatures (above melting) and 
extrapolate this form to the melting temperature in order to obtain $\eta'(T'=T_{melt})$. As we will report elsewhere \cite{nick2}, the theoretically predicted Eq. (\ref{boundratio}) is indeed nearly always satisfied for all types of glass formers examined (a subtle issue is how to precisely extrapolate the 
high temperature viscosity). The upper bound of Eq. (\ref{boundratio}), i.e., a ratio of two (expected
when the latent heat (${\cal{PT}}E{\cal{I}}$) contributions to the average long time drift velocity of Eq. (\ref{stokes}) are very modest) seems to appear most often in fragile glass formers \cite{nick2,ptei--}.

In Fig. \ref{OTP.}b, we plot Eq. (\ref{geoni}) with {\it{a \underline{single} fitting parameter}} $\overline{A} \ll 1$ (consistent with our assumption of ${\overline{\sigma}}_T  \ll T_{melt}$). In the Gaussian approximation to ${\it{p}}_{T}$, we may determine the $T$ dependence of ${\overline{\sigma}}_T$ by equating the numerical value of $\eta_{s.c.}(T)$ in measured data (Fig. \ref{OTP.}a) with the theoretical prediction of Eq. (\ref{geoni}). As implied by a nearly constant value of the dimensionless parameter $A$ in Eq. (\ref{wide'}), the linearity of ${\overline{\sigma}}_T$ in $T$ is evident in Fig. \ref{OTP.}c. Since
$erfc(z \gg 1) \sim \frac{e^{-z^{2}}} {z \sqrt{\pi}}$, we find that
when $(T_{melt} - T) \gg \overline{\sigma}_T= \overline{A} T$, the viscosity
\begin{eqnarray}
\label{geoni+}
\eta_{s.c.}(T) \sim \sqrt{\frac{2}{\pi}}~
{ \overline{\sigma}}_T~ \frac{e^{\frac{(T_{melt}-T)^{2}}{(2 \overline{\sigma}_T^{2})}}}
{(T_{melt}-T)} \eta_{s.c.}(T_{melt}) \nonumber
\\ ={\overline{A}} \sqrt{\frac{2}{\pi}}~\frac{T}{T_{melt}-T} e^{\frac{(T_{melt}-T)^{2}}{2 ({\overline{A}} T)^{2}}}  \eta_{s.c.}(T_{melt}).
\end{eqnarray}
Thus, asymptotically at temperatures far below melting, $\eta_{s.c.}$ scales exponentially in $(\frac{T_{melt}}{T}-1)^2$ similar to fits in \cite{elm}. We derived Eqs. (\ref{geoni}, \ref{geoni+}) having only the single parameter 
${\overline{A}}$. Applying the above calculations {\it mutatis muntandis}, we expect that, similar to viscosity, the scales $\tau_{s.c.}$ setting the long time relaxation rates of other response functions (such as the dielectric response discussed in Section \ref{relaxing}) may be given by Eq. (\ref{geoni}) following the substitution $\eta_{s.c.}(T) \to \tau_{s.c.}(T)$ \cite{achtung}.

\section{Cooperative effects and estimate of the single parameter in the fit.}

The dimensionless parameter $\overline{A}$ measures the amount by which the system deviates from an equilibrated one. 
If $\overline{A} =0$ for states $\{| \psi \rangle\}$ then the system cannot supercool to form a glass by the protocol used
to generate the states $\{| \psi \rangle\}$. In this section, we suggest simple estimates beyond the above trivial statement. 

 As we argue next, our theory suggests that there exists a temperature $T_{A}$, such that for $T<T_{A}$, cooperative effects onset and the liquid {\it may violate the Stokes-Einstein and other relations that hold for thermal systems.} As we noted earlier, at low $T'<T_{melt}$ an equilibrated solid is no longer fully ergodic. Instead, the solid exhibits ergodicity in disjoint phase space regions or the earlier noted ``ergodic sectors'' \cite{nigel}. In that regard, the system is reminiscent of localized systems below the ``{\it mobility edge}'' \cite{Mobility_Edge'} that is set, in our case, by the melting energy density. As the distribution ${\it{p}}_{T}(E')$ may generally have a significant spread of energy densities, it follows that already at some temperature $T_{A}>T_{melt}$, the distribution ${\it{p}}_{T}(E')$ may start to have substantial weight associated with the solid-like states. More generally, a finite relative width of the energy density distribution ${\it{p}}_{T}(E)$ indicates a deviation from the equilibrium result of Eq. (\ref{eq_ratio}) yet that deviation might not be obvious if the equilibrium observables do not change significantly over this energy range. The variation may be pronounced once low energy solid-like states appear. For a Gaussian ${\it{p}}_{T}(E')$ such as that discussed earlier, $(T_{A} - T_{melt}) = {\cal{O}} ({\overline{\sigma}}_T)$ with ${\overline{\sigma}}_T$ the corresponding width in temperatures previously introduced. Indeed, concurrent numerical simulations of metallic glass formers  \cite{numerics',egami} indicate the breakdown of the Stokes-Einstein relation at temperatures only slightly above that of melting. Interestingly, a low temperature deviation from an Arrhenius behavior for the viscosity typical of equilibrated liquids above melting \cite{numerics'} onsets at the same temperature $T_{A}$ at which the Stokes-Einstein relation is violated. Within the framework of the current theory, a larger than Arrhenius (super-Arrhenius) viscosity must appear when ${\it{p}}_{T}(E')$ includes low-temperature solid type states (as indeed occurs below the same temperature $T_{A}>T_{melt}$) and/or liquid type eigenstates that are not ergodic and may thus break the Stokes-Einstein relation. In the viscosity fits displayed in \cite{avoided1} for multiple types of glass formers, a deviation from Arrhenius behavior appeared at a temperature that, on average, for the twenty glass formers investigated therein, was $1.096$ times higher than the melting temperature \cite{avoided2}. A similar average value of $(T_{A} - T_{melt})/T_{melt}$ ($\approx 0.075$) was found \cite{long} for the 23 metallic glass formers analyzed in \cite{metallic-glass}. We speculate that for each liquid, this scale may coincide with $\overline{A}$. Indeed, for the glass forming liquid o-terphenyl (see Fig. \ref{OTP.}), we found when contrasting the simple theoretical prediction of Eq. (\ref{geoni}) with data, a value $\overline{A}= 0.049$ while, experimentally, deviations from bare Arrhenius behavior are indeed already observed at a temperature $0.057$ higher than melting \cite{avoided1,avoided2}. Given the above, we may estimate the single parameter $\overline{A}$ in Eq. (\ref{geoni}). Specifically, at $T_{A}$ the Gaussian ${\it{p}}_{T}(E')$ extends into energy densities below that of melting suggesting that 
 \begin{eqnarray}
\label{Aeq}
\overline{A} \approx \frac{T_{A} - T_{melt}}{T_{melt}}.
\end{eqnarray}
As we remarked earlier, in all liquids that we examined \cite{nick,nick'}, $ \overline{A} \approx 0.05-0.15$. If Eq. (\ref{Aeq}) is correct then it enables an estimate of the single unknown parameter
in our expression for the viscosity below melting (Eq. (\ref{geoni})) from viscosity measurements at high $T$. For completeness, we remark that another possibility is that $T_A$ marks a genuine property of the equilibrated liquid and is not, at all, an outcome of supercooling. 
As noted, at high enough temperatures, the system is thermal. At these temperatures, the distribution ${\it{p}}_T$ must become narrow and cannot be described a Gaussian of finite relative width set by the constant non-thermalization parameter $\overline{A}$. The specific heat of the supercooled liquid is very close to that of the annealed fluid at temperatures above melting.

 \section{Dynamical heterogeneities.} 
 \label{sec:dh}
 As Eq. (\ref{dec}) and characteristic {\bf{(i)}} highlight, non-thermalization implies that {\it there must be a distribution of
 energies} in the relevant eigenstates that comprise the supercooled liquid. 
 As discussed in Section \ref{sec:prob}, the width (as seen by the ratio of Eq. (\ref{wide'})) of the energy densities of the eigenstates appearing in Eq. (\ref{dec}) implies the existence of uncertainties in particle positions and/or momenta. (The proof of this assertion is trivial- if no uncertainty exists then the standard deviation the energy of Eq. (\ref{atomic}) as measured in the state of Eq. (\ref{dec}) will have a vanishing energy density distribution ratio of Eq. (\ref{wide'})). When present, this uncertainty does not imply that the fluctuations of the kinetic energies of spatially far separated particles about their mean values are correlated. Rather, it indicates that the long time temporal fluctuations of the kinetic energies may be nontrivial. That is, if we set the observable  in Eqs. (\ref{llong}, \ref{thermog}) to be the kinetic energy of an individual particle ${\cal{O}}_{1} = \frac{P_{i}^{2}}{2M}$ and squared kinetic energy
 ${\cal{O}}_{2} =  (\frac{P_{i}^{2}}{2M})^{2}$ then if probability distribution ${\it{p}}_{T}(E')$ associated with the state $| \psi \rangle$ would, ideally, not change with time then the variance ${\cal{O}}_{2} - ({\cal{O}}_{1})^{2}$ may be significant (i.e., of the order of the long time average of ${\cal{O}}_{1}$
 itself). An uncertainty in particle momenta (i.e., a broad distribution of their values) naturally relates the appearance of non-uniform local motions
 (``dynamical heterogeneities'' \cite{DH}) that we next further comment on.

 We may associate an effective temperature $T' = U^{-1}(E')$  to eigenstates of energy $E'$.
 Here, $U^{-1}$ is the inverse function associated with the internal energy  $U(T') \equiv E'(T')$ at temperature $T'$ of the equilibrium system defined by Eq. (\ref{atomic}). 
 The mixture of effective temperatures $T'$ (associated with the states appearing in Eq. (\ref{dec})) each with its own characteristic dynamics may lead
 to spatially non-uniform motion. For exponentially activated and other common dynamics, in the equilibrated solid, the local instantaneous relaxation rates differ more substantially at low $T'$. By contrast, at high temperature, small variations in the effective temperatures will not trigger significant change in the dynamics. We may thus expect, the dynamics to be more spatially heterogeneous at low temperature $T$ of the supercooled
 liquid as has been experimentally observed \cite{DH}. 
  
\section{Viscosity above the melting temperature.}
\label{vmt}
In equilibrated liquids, the relaxation rate is given by Eq. (\ref{rrate}). We may apply Eqs. (\ref{stokes}, \ref{fix}) for liquids supercooled to a final temperature $T \gtrsim T_{A}>T_{melt}$. 
In this case, we only have liquid like eigenstates in the decomposition of Eq. (\ref{dec})
and the lower $T'> T_{melt}$ cutoff in the integrals of Eqs. (\ref{stokes}, \ref{fix}) becomes essentially irrelevant. Because typically, 
the Gibbs free energy does not vary strongly with $T'$ for temperatures close to ``$T_{A}$'', 
the viscosity of the supercooled liquid may be approximated by an Arrhenius form with an activation barrier $\Delta G$, i.e., the {\it equilibrium viscosity} is
$\eta(T') = \eta_{0} \exp(\Delta G(T')/(k_{B} T'))$ with a constant $\eta_{0}$ and nearly constant $\Delta G(T')$. 
By Eqs. (\ref{stokes},\ref{fix},\ref{rrate}), if ${\cal{PT}}E{\cal{I}}$ contributions may be ignored then
\begin{eqnarray}
\label{esc}
\eta_{s.c.}(T) = \frac{\eta_{0} }{\int_{T_{melti}}^{\infty}  dT'~ C_{v}(T')~e^{-\frac{\Delta G(T')}{k_{B} T'}} {\it{p}}_{T}(E'(T'))}.
\end{eqnarray} 
Elsewhere, it was illustrated (both theoretically (for general liquids using WKB \cite{long}) and experimentally (for metallic glass formers \cite{metallic-glass})) 
that $\eta_{0} \simeq nh$ with $n$ the average spatial particle number density and $h$ Planck's constant.
Since $\Delta S = - \frac{\partial}{\partial T} \Delta G$ \cite{long},
\begin{eqnarray}
\label{dS}
\Delta S = - \frac{\partial}{\partial T} (k_{B} T \ln \eta(T)),
\end{eqnarray} 
while the energy (or, more precisely, enthalpy) barrier is 
\begin{eqnarray}
\Delta {\sf H} = k_{B} T \ln (\frac{\eta}{\eta_{0}}) - T  \frac{\partial}{\partial T} (k_{B} T \ln \eta(T)).
\end{eqnarray}
The entropy difference $\Delta S$ associated with enthalpy barriers $\Delta {\sf H}$ may be probed by invoking the measured $T$ dependence of the viscosity.
A nearly constant free energy activation barrier implies that $\Delta S \sim 0$. The temperature $T_A$ may be found by setting Eq. (\ref{dS}) to zero. For {\it equilibrated liquids} \cite{long}, $\Delta S \ge 0$.

\begin{figure*}
\centering
\includegraphics[width=2.15 \columnwidth]{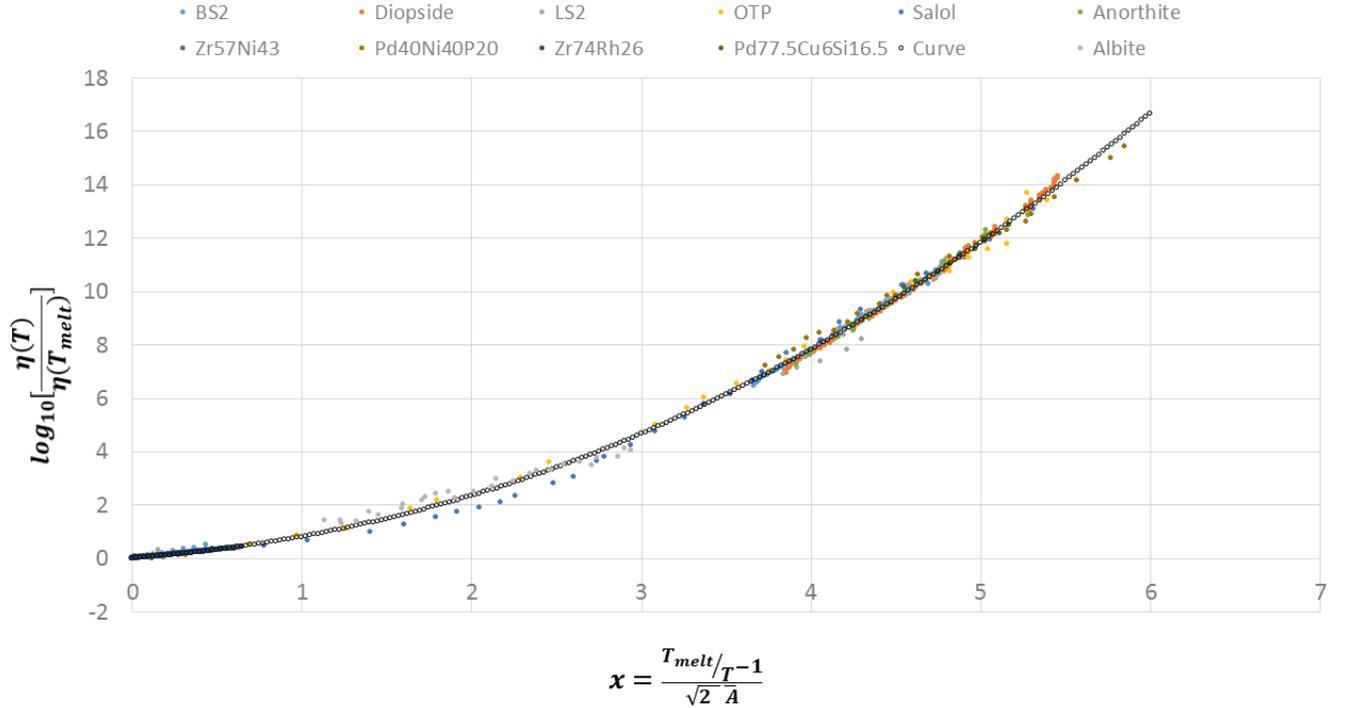}
\caption{(Color online.) From Weingartner et al., \cite{nick}. The collapse predicted by Eq. (\ref{geoni})  (``curve'' in the above legend) \cite{me1}
is shown here for liquids that span diverse types of glass formers (organic, metallic, and silicate). 
We plot the logarithm of the viscosity of divided by the viscosity at the melting temperature
as a function of a ``reduced'' temperature $(T_{melt} - T)/ (\overline{A}T \sqrt{2})$ (with the parameter $\overline{A}$ and experimentally measured melting temperature $T_{melt}$ associated with each of the respective fluids), see text. The viscosity collapses over 16 decades of data. }
\label{nussinov_collapse}
\end{figure*}

\section{Predictions and conclusions.} Using elementary statistical mechanics considerations, we developed a simple unified framework for understanding the ``glass transition''. We argued that the observed behaviors of supercooled fluids can be understood in terms of melting transitions that equilibrium systems undergo without needing to invoke any other assumptions concerning additional bona fide transitions. In the current formalism, the non-adiabatic supercooling process merely leads to a broadening of the probability distribution associated with different eigenstates. We suggested that all the behaviors of supercooled liquids and glasses may be understood and quantitatively computed with the aid of this probability distributions. We explained that all ``transitions'' in the supercooled liquid may stem from those exhibited by the equilibrium system at its melting temperature (as well as any possible crossovers or singularities of the probability distribution). When the melting transition is associated with latent heat
(as it typically does), the equilibrium transitions are rounded off at the temperatures of the supercooled fluid that correspond to the energy density interval $E_{melt}^{+}/V \ge E/V \ge E_{melt}^{-}/V$. The energy densities $E_{melt}^{\pm}/V$ mark the boundaries of the liquid-solid coexistence region in the equilibrium system (the energy density domain that we denoted by ${\cal{PT}}E{\cal{I}}$ in the current work). The lower of these energy densities may correspond to a temperature ($T_{-}$, see Section \ref{sec:Energy}) of the supercooled liquid that is far below the melting temperature.

As we described in the Introduction and touched on throughout this work, the quantum formalism was employed as a computational tool.  
Nowhere in our derivation were our final results inherently quantum. Although $\hbar$ initially appeared in the calculations (e.g., Eqs. (\ref{llong}, \ref{lllong})), $\hbar$ dropped out in our final, essentially classical, relations of Eqs. (\ref{thermog}, \ref{richer_dep}). Elsewhere we will show that independent classical considerations suffice for deriving the current results \cite{nick',nick2}. The advantage of the quantum approach is that the requisite pertinent classical ``microstates''  were made precise. Indeed, as we emphasized in the Introduction, even in the standard definitions of the partition function and entropy of
classical systems, quantum mechanics rears its head when counting microstates in an other continuous system by, essentially, introducing Planck's constant to define the minimal volume in phase space. 
Albeit initially appearing in the classical textbook phase space partition function, Planck's constant typically drops out from the final standard classical equilibrium averages that are computed with this partition function. A density matrix describes our system in much the same way as it does any other ``classical'' system.

Fluctuations in the energy density typically tend to zero in the thermodynamic limit. In the current theory we allow for states with differing energy densities. We do so since, as noted in Section \ref{sharp}, if the system is in contact with a heat bath then within our approach (as in all others) there will be no fluctuations in the energy density. Barring special quantum effects (see Appendix \ref{mbl-a}), in order to
for the system to be out of equilibrium (see item {\bf{(i)}} of Section \ref{sec:prob}), the density matrix must, however, contain states that are of differing energy density (i.e., the dimensionless ratio $\overline{A}$ of Eq. (\ref{wide'}) cannot vanish). To further illustrate this point, general considerations and simple toy calculations 
were advanced in Sections \ref{sec:non-locaiity} and \ref{e-fluc}. The appearance of states of varying energy densities in the density matrix (as mandated by our theory) is pleasing as it naturally predicts the existence of the experimentally observed features such as dynamical heterogeneities (see Sections \ref{nontrivfinal} and \ref{sec:dh}). More broadly, qualitative analogs of a finite ${\overline{A}}$ (for quantities other than the energy) appear in 
various random systems, e.g., \cite{amnon,per} that are known to exhibit ``non-self-averaging'' in their thermodynamic limit.

The central prediction of our approach is that the experimental thermodynamic and dynamic measures of supercooled fluids will, respectively, 
be given by Eqs. (\ref{constraint},\ref{thermog}) and Eqs. (\ref{stokes}, \ref{fix}). In these equations, all quantities are measurable for the equilibrated system 
apart from the probability density ${\it{p}}_{T}(E')$. This density is constrained by Eq. (\ref{constraint}). We argued that the probability distribution for the energy density may be a simple Gaussian, Eqs. (\ref{wide'},\ref{Gauss''}). This probability distribution leads to a very simple expression for the predicted viscosity of supercooled liquids (Eq. (\ref{geoni})) below the melting (or liquidus) temperature. As a proof of principle of our simple ideas, in Fig. \ref{OTP.}b we fitted viscosity data of a glass forming liquid with the single parameter ${\overline{A}}$ of Eq. (\ref{wide'}) potentially linked with a
deviation from an Arrhenius behavior. In companion works \cite{nick,nick'}, we will report that the single parameter in our theory is nearly constant across many different glass formers, $\overline{A} \sim 0.05-0.15$. If correct, Eq. (\ref{geoni}) requires that viscosity data from different glass formers may be made to {\it universally collapse with the use of the single dimensionless parameter} ${\overline{A}}$. 
In particular, if the dimensionless scaled viscosity data $\eta/\eta(T_{melt})$ is plotted for different fluids as a function of the dimensionless ``reduced'' temperature $(T_{melt} - T)/ (\overline{A}T \sqrt{2})$, the data should be made to collapse with the single adjustable parameter $\overline{A}$. In \cite{nick,nick'}, we demonstrate that this indeed exactly occurs for many different
types of glass formers including silicates, metallic glass formers, and organic fluids. Figure \ref{nussinov_collapse} demonstrates this collapse. A further consequence of our theory for the viscosity is 
the bound of Eq. (\ref{boundratio}). In future work \cite{nick2}, we will report that this inequality is indeed satisfied with the upper bound being saturated in fragile glassformers. Clearly, although yielding reasonable agreement with the experimental data
(as evidenced in Fig. \ref{eigen.}b, and the above noted collapse of Fig. \ref{nussinov_collapse}), the Gaussian form of ${\it{p}}_{T}(E')$ motivated by Maximum Entropy considerations in which the temperature $T$ provided the only energy scale need not generally hold. In order to capture the experimentally seen peaks in the heat capacity at $T_{g}$ and other detailed phenomena, specific temperatures (or associated energies) scales must appear in the distribution ${\it{p}}_{T}$. In this work, we adopted an Occam's razor approach wherein the smallest number of parameters (i.e., the single dimensionless parameter $\overline{A}$) were invoked in this unknown distribution. In Appendix \ref{table_ar}, we provide a table summarizing all of the above that highlights the key assumptions made in our work and the results that they led to. 

The quantities in Eqs. (\ref{thermog}, \ref{fix}) may be varied by applying external magnetic fields in metallic glass systems or external strains on more general systems; these will then trigger changes in measured thermodynamic and dynamic response as predicted by these equations. 
Each measurement of a different observable ${\it{O}}$ or a relaxation rate will lead to an additional independent constraint on ${\it{p}}_{T}(E')$.
By choosing the functions ${\cal{O}}$ in Eqs. (\ref{thermog}) to, e.g., be the Hamiltonian itself or the local particle density ${\sf n}'(\vec{x})$, we may respectively relate the energy (and specific heat) or the local structure of the equilibrium system to that of the supercooled liquid. We explained how the amorphous structure of glasses and dynamical heterogeneities arise naturally within our framework. In Appendix \ref{space.}, we further suggest how low energy {\it locally preferred structures may arise naturally}. Such structures have been predicted and experimentally observed \cite{LPS} and have long underlined many theories \cite{avoided1,avoided2,avoided3,LPS-theory}. 

Another {\it prediction} of our theory is that
the photon {\it emission spectrum} be related to that of equilibrium systems at different energy densities $E'/V$, or associated temperatures $T'$, when weighted by the distribution ${\it{p}}_{T}(E')$. Specifically, the energy density $u_{s.c.,~\omega}(T)$ associated with photons of a given frequency 
$\omega$ 
emitted by the supercooled liquid is given by 
\begin{eqnarray} 
\label{usc}
u_{s.c.,~\omega}(T) = \int dE' {\it{p}}_{T}(E') u'_{\omega}(E').
\end{eqnarray}
Here, $ u'_{\omega}(E')$ is the energy density carried by photons of frequency $\omega$ when these are in equilibrium with a thermal system of energy density $E'/V$ (or, equivalently, of temperature $T'$ at which its energy density is the internal energy density of the equilibrated system).
This suggests that the emission spectrum of a supercooled liquid having a temperature $T$ may be related to that in the equilibrium solid via a smearing given by the 
distribution ${\it{p}}_{T}$.

Although we have focused on structural glasses, the results of Eqs. (\ref{constraint},\ref{thermog}) are general and may be applied to other, very different, systems. For instance, one may consider metals given by Eq. (\ref{atomic}) (or a simplification thereof)
augmented by an applied external electric field for which the charge current
\begin{eqnarray}
\label{jv}
\vec{J} = \sum_{j} q_{j} \vec{v}_{j}.
\end{eqnarray} 
Here, $\{q_{j}\}$ are the charges of the carriers and $\{\vec{v}_{j}\}$ their corresponding velocities. Similar to Eq. (\ref{stokes}), one may compute both the long time average of the current $\vec{J}$ (and thus the DC conductivity) as well as thermodynamic observables (e.g., Eq. (\ref{constraint})) from the probability distribution ${\it{p}}_{T}$. Our framework may  therefore be of use in inhomogeneous electronic systems, e.g., \cite{davis2}. 

Eqs. (\ref{nET},\ref{o-average}) hold for general fluctuations in the particle number density or chemical potential (with these augmenting the fluctuations in the energy density or effective equilibrium temperature that we largely focused on in the current work). One may superpose states of different particle densities in Eq. (\ref{dec}). In such a case, the resultant probability distribution ${\it{p}}$ is a function of both an effective temperature $T'$ and a chemical potential $\mu'$ of the equilibrium system. Superposing individual states with varying carrier or other particle number densities might, e.g., lead to a smearing of the Fermi surface and ensuing deviations from Fermi liquid type behaviors in electronic and other systems. 

For completeness, we comment on another route towards non-ergodicity. Such a path is afforded by not only having an energy density distribution of a finite width but by being constrained to a subspace of the set of all states of a given energy density.
If and when this occurs, such a confinement to a constricted subspace may impact many properties that are not solely energy dependent (possibly including structure). As will be detailed in Appendix \ref{mbl-a}, the probability for that to occur is low; we will further discuss a rigorous monotonicity property that constrains the appearance of such localized states. However, even if such a confinement does transpire, if the property computed depends only on the energy then being constrained into a subspace does not matter- all that
is important is the energy alone. In essence, this is our assumption for the viscosity and other hydrodynamic properties.

Our theory relies on the behaviors of the equilibrated system. Hence, if the equilibrium transitions in systems with disparate local or long-ranged Hamiltonians $H$ are notably distinct from one another then it is natural to anticipate different dynamics in these systems upon supercooled. Similarly, the equilibrium (and non-equilibrium) fluctuations in local systems may be different from those in systems with long range interactions \cite{brief}. We speculate that this may reconcile numerical observations concerning the seeming lack of correlation between structure and dynamics, e.g.,  \cite{structure_gilles}.  

We close with general remarks. The existence of the single parameter form for the viscosity may enable a simplified practical analysis of glasses and supercooled liquids. In \cite{nick}, we will demonstrate the empirical validity of Eq. (\ref{geoni}) for all tested supercooled fluids. It may be that our prediction for the viscosity works well for serendipitous reasons independent of the guiding theoretical principles that led us to suggest Eq. (\ref{geoni}) and the associated universal collapse that it mandates. As we hinted earlier, beyond the specific functional form of Eq. (\ref{geoni}), the possibility of such a collapse was already suggested by the exact relation of Eq. (\ref{stokes}). In the works of \cite{vft1} and others since, no theory nor phenomenological fit was found where a single dimensionless parameter describes the viscosity of all glass formers. 
In the appendices, we summarize the assumptions that we have made and their implications (Appendix \ref{table_ar}),
qualitatively sketch how the concepts advanced 
in this work may naturally lead to the empirical facts underlying the Kauzmann paradox (Appendix \ref{walter-sec}), relate our results to the spatial structure (Appendix \ref{space.}), expand on the earlier noted possibility of many body localized eigenstates (Appendix \ref{mbl-a}), comment on connections to low temperature glasses (Appendix \ref{low-sec}), and discuss in Appendix \ref{multi-sec}
the possibility of having multiple temperature or energy scales present in the distribution ${\it{p}}_{T}(E')$ (i.e., additional scales aside from those set by the temperature $T$ itself).

%A natural test of the theory is to see whether such distributions simultaneously lead to the correct prediction for both the static and dynamic quantities (Eqs. (\ref{thermog}, \ref{fix})).

{\it{Acknowledgements}}. I am extremely grateful to N. Weingartner for help with the numerical fits and gratefully acknowledge work with N. B. Weingartner, F. S. Nogueira, K. F. Kelton, M. Blodgett, E. Altman, S. Banerjee, and L. Rademaker on related problems. In particular, in \cite{nick,nick',nick2} we tested and extended many of the ideas that were initially introduced here. I wish to thank G. Biroli, G. Parisi, and especially to T. Schaefer for prompting me to write up my approach and am appreciative of positive remarks by T. Egami and questions by F. Zamponi. This research was supported by the NSF DMR-140911229 and the Feinberg foundation visiting faculty program at Weizmann Institute.

\appendix

 \clearpage
\section{Table of assumptions and results}
\label{table_ar}

In the table below, we sketch the key assumptions that we made in the current work and the results that they implied. 

\begin{center}
\begin{tabular}{ | m{21.5em} | m{21.5em}| } 
\hline
Assumptions  & Results \\ 
\hline
{\bf{(0)}}  Equilibrium statistical mechanics may be applied to generic systems that, in the absence of supercooling, are experimentally known to display equilibrium liquid and solid phases. In particular, the measured equilibrium quantities are equal to their microcanonical ensemble averages. & As a function of the energy density, the eigenstate average exhibits non-analyticities at the transition points
(see conclusions {\bf{(A)}} and {\bf{(B)}} of Section \ref{generalc}). \\
\hline
{\bf{(1)}} Invoked the idealized supercooling protocol of Eq. (\ref{HHH}). & The probability density ${\it{p}}_{T}(E')$ of Eq. (\ref{ptttt}) is time independent. \\
\hline
{\bf{(2)}} Assumed that, when the observables ${\cal{O}}$ of interest are projected to constant energy subspaces
(as embodied by the operators $[\delta (E'-H) ~{\cal{O}} ~\delta (E'-H)]$), the eigenvalues are regular functions of the energy $E'$ 
and any other quantum numbers ${\cal{Q}}$ on which they may depend. &
Eq. (\ref{richer_dep}) (and its generalization of Eq. (\ref{o-average})) relates the long time average of any observable to weighted
averages over equilibrium expectation values in different ergodic sectors (quantum numbers) ${\cal{Q}}$. \\
\hline
{\bf{(3)}}  Ignored, in Eq. (\ref{richer_dep}), a possible dependence on special quantum numbers ${\cal{Q}}$ other than energy. & 
When, in {\it equilibrated systems}, a dependence on quantum numbers ${\cal{Q}}$ is indeed not present for the observables ${\cal{O}}$ of interest then 
Eq. (\ref{thermog}) may be rigorously employed.\\
\hline
{\bf{(4)}} Using information theoretic arguments, we motivated yet, nevertheless, still assumed that, similar to equilibrium statistical mechanics, a Gaussian distribution ${\it{p}}_{T}(E')$ appears for the energy (Eq. (\ref{Gauss''})). &
{\bf{(i)}} This assumption enabled us to substitute a Gaussian ${\it{p}}_{T}(E')$ form in all pertinent integrals stemming from Eq. (\ref{thermog},\ref{shorteq}).
{\bf{(ii)}} To avoid equilibration, the inequality of Eq. (\ref{wide'}) needs to be satisfied (a {\it finite} dimensionless 
parameter $\overline{A}$ must be present). \\
\hline
{\bf{(5)}} The dimensionless parameter $\overline{A}$ defined by Eq. (\ref{wide'}) is small ($\overline{A} \ll 1$) and constant for the temperatures $T$ 
for which the viscosity is experimentally measured,
$T_{melt} \ge T \ge T_{g}$. The smallness of $\overline{A}$ amounts to an assumption that, albeit its finite width, the energy distribution is not radically different from that in the equilibrium system. 
& For the implied rapid drop of the distribution ${\it{p}}_{T}(E')$ with the energy $E' > E_{melt}$ for temperatures below melting, $T<T_{melt}$, we arrive at the inequality of Eq. (\ref{vis}) 
relating the distribution ${\it{p}}_{T}(E')$ to the viscosity.  \\
\hline
{\bf{(6)}} 
Posited that within the  ${\cal{PT}}E{\cal{I}}$ there are, effectively, no contributions to the long time average velocity $v_{l.t.a.}$. In other words, we assumed that in equilibrium within the  ${\cal{PT}}E{\cal{I}}$  region the terminal velocity $v'_{\infty}$ is insignificant. 
& The inequality of Eq. (\ref{vis}) became an equality. In particular, combining all of the above assumptions, we obtained our central prediction of 
Eq. (\ref{geoni}) for the viscosity as a function of temperature. \\
\hline
\end{tabular}
\end{center}
\clearpage

 \section{Hallmarks of the Kauzmann paradox.} 
 \label{walter-sec}
 Eqs. (\ref{thermog},\ref{stokes},\ref{fix}) tie both the thermodynamics and dynamics within supercooled liquids to a single object- the probability density ${\it{p}}_{T}(E')$. As we now describe, this dependence seems to {\it qualitatively} give rise to the phenomenology inderlying the ``Kauzmann paradox'' \cite{`walter}. The crux of this paradox is that when the measured entropy of a supercooled liquid is extrapolated to low temperatures, it nearly coincides with that of the equilibrium solid at the (Kauzmann) temperature $T=T_{K}$. At yet lower temperatures, $T<T_{K}$, the extrapolation will lead to the paradoxical situation that the entropy of the supercooled liquid would be lower than that of the equilibrated solid. Empirically, $T_{K}$ is close to $T_{0}$ of the VFTH viscosity fit. A commonly held viewpoint is that an ``ideal glass'' transition intervenes as $T$ is lowered (leading to a divergent relaxation time) thus resolving the paradox for the unaccessible temperatures $T<T_{K}$. 
 
 Our framework specifically rationalizes the essential phenomenology without assuming an ideal glass transition. Suppose (as consistent with the constraint of Eq. (\ref{constraint})) that seemingly, by extrapolation, the integral $\lim_{T \to T_0} \int_{E_{melt}}^{\infty} ~{\it{p}}_{T}(E')~ dE' = 0$ then, by Eqs. (\ref{stokes}, \ref{fix}), $\lim_{T \to T_{0}} r_{s.c.}(T ) =0$. Consequently, only within the above noted {\it extrapolation}, $\lim_{T \to T_{0}}\eta_{s.c.}(T) = \infty$. On the other hand, from Eq. (\ref{vis}), when the above extrapolated integral vanishes, the long time average of a thermodynamic quantity ${\cal{O}}$ in the supercooled liquid must coincide with its average over equilibrated  low-temperature solid-like states. This is so as ${\it{p}}_{T}(E')$ has all of its support from the solid like states once the above integral vanishes. With Eq. (\ref{thermog}), we may calculate the internal energy and general thermodynamic functions. Integration of ($C_{p}(T)/T$) yields the entropy. Thus, the entropy (similar to other measures) of the supercooled liquid may approach that of the solid when $T \to T_{0}$. This suggests that $T_K \simeq T_0$ yet not necessarily as a precise equality as the phenomenon concerns a low $T$ extrapolation of ${\it{p}}_{T}(E')$ as it appears in Eqs. (\ref{thermog}, \ref{stokes}, \ref{vis}). Experimentally, for different liquids, $(T_K/T_0)$ is, indeed, not exactly unity but rather lies in the range $0.9-1.1$ \cite{t0tk}. We emphasize that within our approach, the viscosity does not diverge at any positive temperature.

\section{Spatial structure.} 
\label{space.}

As we explained in Section \ref{sec:prob}, the significant width (as evinced by the ratio of Eq. (\ref{wide'})) of the energy distribution mandates an uncertainty in particle positions and/or momenta (as if no uncertainty existed, the Hamiltonian of Eq. (\ref{atomic}) would have no uncertainty as well and thus the width $\sigma_T$ would need to vanish). Thus, long range order might not, generally, appear in the supercooled liquid (and glass)- as it indeed does not. We now further discuss the absence of ordering from complementary perspectives.

From Eq. (\ref{constraint}) and the positivity of the probability density, $E(T) \ge E_{melt} \int_{E_{melt}}^{\infty} {\it{p}}_{T}(E') dE' + E'(0) \int_{E'(0)}^{E_{melt}}  {\it{p}}_{T}(E') dE' =
E_{melt} \int_{E_{melt}}^{\infty} {\it{p}}_{T}(E') dE' +  E'(0) (1- \int_{E_{melt}}^{\infty} {\it{p}}_{T}(E') dE' )$. Here, $E'(0)$ is the ground state energy (the zero temperature energy of the equilibrated system) which we define to be the zero of the energy, $E'(0)=0$. Thus, {\it rigorously}, at sufficiently low $T$, the cumulative probability of being in any liquid state, 
\begin{eqnarray}
\label{liquid_prob}
P_{liquid} \equiv \int_{E_{melt}}^{\infty} dE~ {\it{p_{T}}}(E)
 \le  \frac{E(T)}{E_{melt}}.
 \end{eqnarray}
If the energy density of the supercooled liquid at low enough $T$ is strictly bounded from above then given Eq. (\ref{liquid_prob}), $P_{liquid}$ can be made small. Any system, including an amorphous glass, of sufficiently low energy $E(T)$ must mostly consist of {\bf (1)} {\it solid like} eigenstates when decomposed as in Eq. (\ref{dec}) and/or of ({\bf{2}}) {\it mixed liquid-solid} type eigenstates that lie in the latent heat energy interval (the shaded ${\cal{PT}}E{\cal{I}}$ region of Fig. \ref{eigen.}). From Eq. (\ref{liquid_prob}) and the experimental fact that low $T$ glasses do not exhibit crystalline structure, it is evident that a high probability $(1-P_{liquid})$ of being in solid or ${\cal{PT}}E{\cal{I}}$ eigenstates {\it does not} imply that that the system has crystalline order. The scattering intensity associated with momentum transfer  $\Delta \vec{p} = \hbar \vec{k}$ is given by the structure factor $S_{\vec{k}} = |\int d^{d}x ~e^{i \vec{k} \cdot \vec{x}} \langle {\sf n}(\vec{x}) \rangle|^2 \equiv | \langle {\sf{n}}_{\vec{k}} \rangle|^2$
with $\langle {\sf n}(\vec{x}) \rangle$ the spatial density of the scatterer and ${\sf{n}}_{\vec{k}}$ its Fourier transform.
We briefly expand on contributions of type {\bf (1)} - those of solid type eigenstates in systems that, in equilibrium, display low temperature crystalline order.  Such solid type eigenstates differ by (a) symmetry operations such as rotations and spatial shifts ($\vec{x} \to \vec{x} + \vec{a}$ with any displacement $\vec{a}$ restricted to the unit cell) and, notably, (b) in their individual energy densities (associated with the spread of energies
in ${\it{p}}_{T}(E')$). (a) and (b) together with the contributions of ${\cal{PT}}E{\cal{I}}$
eigenstates  may lead to a heterogeneous $\langle {\sf n}(\vec{x}) \rangle$. Thus, even without the ${\cal{PT}}E{\cal{I}}$ contributions, the bound of Eq. (\ref{liquid_prob}) may be satisfied (as it must) with a small $P_{liquid}$ in an amorphous supercooled system. As we elaborated in the main text, not only is the structure non-trivial but also other counterintuitive properties appear in systems with a broad
energy density distribution (in which, as we briefly reviewed above, positional and/or momenta uncertainties must appear). While single crystals as well as powder samples of crystallites (large on an atomic scale such that each crystallite on its own leads to a finite $S_{\vec{k}}$ only if $\vec{k} = \vec{K}$ where $\vec{K}$ is a reciprocal lattice vector) lead to sharp diffraction patterns, Eq. (\ref{liquid_prob}) implies that low energy states need not display crystalline structure since the amplitudes $\{c_{n}\}$ in Eq. (\ref{dec}) may span many distinct eigenstates. Regarding (a) above, unlike the uniform sign observables that we largely focus on in the main text that are smooth single valued functions of the energy, the density may vary in degenerate eigenstates. That is, one cannot determine the expectation value of the density $\langle {\sf n} (\vec{x}) \rangle$ at a specific point $\vec{x}$ in a crystal given only its temperature (or energy density). The low temperature solid like eigenstates splinter into multiple ``ergodic sectors'';  each of these sectors has a different value of $\langle {\sf n}(\vec{x}) \rangle$ \cite{nigel}. Contrary to typical thermal states with spontaneously broken symmetries, the superposition in Eq. (\ref{dec}) 
mixes degenerate states. The different phases of the contributions from the different eigenstates in $S_{\vec{k}}$ can lead to interference effects.
Furthermore, mixed liquid-solid type eigenstates from the ${\cal{PT}}E{\cal{I}}$ lead to contributions that augment those from the solid like eigenstates. 
Within the ${\cal{PT}}E{\cal{I}}$ melt of equilibrium crystals, one often finds nucleated low energy local structures of various sorts. This implies that the corresponding supercooled liquids may {\it exhibit locally preferred structures} \cite{LPS}. Given the superposition in Eq. (\ref{dec}), structure computed from the real space density matrix or structure factor measurements of the supercooled liquid (containing solid like and mixed liquid-solid type eigenstates) will generally differ from that of the low $T$ equilibrated solid (associated with eigenstates over a vanishing energy density interval).  

We may formalize the above considerations by employing Eq. (\ref{richer_dep}) with ${\cal{O}} ={\sf{n}}_{\vec{k}}$. We may write this equation anew and explicitly apply it to the structure factor,
\begin{eqnarray}
\label{richer_dep+}
 ({\sf n}_{\vec{k}})_{l.t.a.} &&= \sum_{\cal{Q}} \int_{0}^{\infty} dE'~  {\it{p_{T}}} (E'; {\cal{Q}}) ~ {\sf n}_{\vec{k}} (E'; {\cal{Q}}) \nonumber
\\ &&= \sum_{\cal{Q}} \int_{0}^{\infty} dT' ~{\sf n}_{\vec{k}}(E'(T'),{\cal{Q}}) ~{\it{p}}_{T} \big(E'(T');{\cal{Q}} \big) ~ C'_{V}(T',{\cal{Q}}) \nonumber
\\ &&+ \sum_{{\cal{Q}}}  \int_{{\cal{PT}}E{\cal{I}}} dE' ~ {\it{p_{T}}} (E';{\cal{Q}}) ~ {\sf n}_{\vec{k}} (E',{\cal{Q}}).
\end{eqnarray}
At low energies, there may be degenerate crystalline eigenstates related by rotational and other symmetries. Thus, in this case, for any wave vector $\vec{k}$,
the ``quantum numbers'' ${\cal{Q}}$ in Eq. (\ref{richer_dep+}) relate to the (different) values that $S_{\vec{k}}$ may assume for eigenstates of the same energy. If the state $| \psi \rangle$ or the density matrix $\rho$ associated with ${\it{p}}_{T}$ contain multiple degenerate energy eigenstates related by symmetry (such that the weight associated with each particular
reciprocal lattice vector ${\vec{K}}$ is small) then the sum (integral) in the second line of Eq. (\ref{richer_dep+}) that runs
over all of these states will lead to a very weak
value of $(S_{\vec{k}= \vec{K}})_{l.t.a.}$. Concretely, one may consider a number density given by ${\sf n}_{\vec{K}}(\vec{x}) = A (1+ \cos(K_{x} x))(1+ \cos K_{y} y) (1+ \cos K_{z} z)$ with a positive constant $A>0$; this describes an orthorhombic  ``crystal'' like density with periodicities $2 \pi/K_{x,y,z}$ along the $x, y$ and $z$ directions. If one uniformly superposes the state associated with $\vec{K}$ with other states generated by rotating $\vec{K}$ on a sphere of radius $K = |\vec{K}|$ assigning, e.g., to each of these individual crystalline states a common squared modulus (and possibly arbitrary phase) then the number density ${\sf{n}}$ of the resultant state (formed by superposing all of these crystalline states) will not exhibit sharp Bragg spots at any particular non-zero wave-vector $\vec{K}$. Rather, the Fourier weights ${\sf n}_{\vec{k}}$ will be smeared for wave-vectors $\vec{k}$ that lie on a sphere of radius $K$. Moreover, as we highlighted above,
there are generally ${\cal{PT}}E{\cal{I}}$ contributions (the last line of Eq. (\ref{richer_dep+})) in which even the equilibrium states do not exhibit any sharp Bragg peaks. 

If there are further notional low energy non-crystalline eigenstates $\{ |s \rangle \}$ of $H$ that are degenerate with crystalline eigenstates $\{|c \rangle\}$, then their dynamics when subjected to a 
perturbation $H_{1}^{u} = \tilde{U'}^{\dagger} H_{1} \tilde{U'}$ will be identical to those of the crystalline states when subjected to the perturbation $H_{1}$. In the above, $\tilde{U'}$ 
is any element of the full symmetry group $G= \otimes_{E} SU(n_{E})$ of the Hamiltonian $H$ linking the two states, $\tilde{U'} | c \rangle = | s \rangle$. Here, $n_{E}$ is the number of eigensates of $H$ of energy $E$ and the product is over all eigenvalues $E$ of $H$. If there is a typical relaxation time of the crystalline states for generic perturbations $H_{1}$ then the same relaxation time appears for the corresponding perturbations $\{H_1^{u}\}$ acting on $| s \rangle$. %If this is true for general perturbations then our results for $ \tau_{s.c.}(T)$ (Section \ref{sec:central}) and other like it will remain unchanged.

We conclude with a further speculation. Recent results \cite{Ankit_Agrawal} illustrate that as the chemical composition of certain metallic glassformers is varied, at conventional cooling rates, these materials go from being glassformers to crystals with the melting temperature of the crystals merging continuously with the glass transition temperature of the glass formers. This finding is consistent with very simple relations: (1) States having an energy density below those of the bottom of the ${\cal{PT}}E{\cal{I}}$ (with these energy densities lying below the energy density at the glass transition temperature) correspond to the solid type eigenstates in equilibrated systems while (2) states in the ${\cal{PT}}E{\cal{I}}$ of the equilibrated system mirror the states found in supercooled liquids above their glass transition temperature.

\section{Many body localized states.}
\label{mbl-a}

As the reader hopefully may have noted, making use of the equilibrium microcanonical ensemble average of Eq. (\ref{mc}) for the disorder free system defined by Eq. (\ref{atomic}) 
(that is known to equilibrate) greatly simplified all of our calculations. We next explicitly ask if our results differ if $H$ has athermal ``many body localized'' eigenstates \cite{MBL1,MBL2,MBL3,MBL4,MBL5,MBL6} for which Eq. (\ref{etheq}) fails. Formally, in such case we may turn to our general relation of Eq. (\ref{o-average}). This latter relation 
holds for both equilibrated as well as many body localized systems. A single many body localized state may be specified by providing explicit quantum numbers ${\cal{Q}}$ that constrain the system to a specific state.

 In the main text, we employed our key relation of Eq. (\ref{thermog}) for the terminal velocity of a sphere dropped into the supercooled liquid. Eq. (\ref{thermog}) is the simpler version of Eq. (\ref{o-average}) that is valid in thermal systems in which the energy density (or temperature) is the dominant variable that governs the observable measured (in our
case, that observable is the vertical velocity). Eq. (\ref{thermog}) does not admit a dependence of this observable on additional quantum numbers. We now return to and motivate why we may employ Eq. (\ref{thermog}) instead
of the more general equality of Eq. (\ref{o-average}). 

As we underscored earlier and just alluded to above, empirical observations attest that both equilibrated liquids and equilibrated solids described by Eq. (\ref{atomic}) thermalize: the micro-canonical ensemble equality of Eq. (\ref{mc}) holds for these systems. That is, for a system size independent large energy interval 
$\Delta E$, all observables satisfy Eq. (\ref{mc}). Many body localized can still exist. However, since the microcanonical ensemble applies to thermal systems (including the standard
disorder free liquids of Eq. (\ref{atomic})),
the average over all eigenstates in energy intervals 
$[E'-\Delta E,E']$ must lead to an equilibrium thermodynamic average. Now, here is an important point that permeates our entire reasoning and that we repeatedly touched on throughout this work: If the probability density ${\it{p}}_{T}(E')$ spans an interval that is extensive in the system size- i.e., it does not correspond to a unique energy density then we may tessellate this extensive energy interval with many segments of width $\Delta E$. In each of these intervals, Eq. (\ref{atomic}) holds and our previous calculations are valid. Typically the spread in energies associated with Eq. (\ref{pt+}) cannot be a finite system independent width $\Delta E$. If the spread in energies were system size independent then the supercooled system (which is out of equilibrium) would satisfy the microcanonical ensemble relation of Eq. (\ref{mc}). However, satisfying Eq. (\ref{mc}) implies that the supercooled liquid is in equilibrium which is not the case. A highly unlikely loophole to this reasoning is still possible: namely, a decomposition of the supercooled state (formed by a Hamiltonian different from $H$) may, miraculously, lead to a pure eigenstate of $H$ or a special sparse set of eigenstates of $H$ which do not cover a finite size energy window yet constitute a large weight of $| \psi \rangle$. For an arbitrary cooling operator $\tilde{U}(t_{final}, t_{initial})$ leading to $| \psi \rangle$, even if Eq. (\ref{atomic}) exhibits many body localized eigenstates (for which the relaxation rates vanish),
 {\it the likelihood of having such a special, highly nonuniform, covering is exceedingly low}. 
 
Bolstering the broad arguments above, a more potent {\it monotonicity} property {\it leads to a constraint} that all states (including putative localized states) must satisfy. This property suggests that, unless they are always degenerate (or very closely so) and appear in unison with non-localized states that enable high enough mobility, no localized states may have an energy density higher than that of eigenstates in which flow can occur. To illustrate, we return to the example of the dropped sphere of Section \ref{vlv} of the main text. If the viscosity of the liquid is monotonically decreasing with increasing temperature (as it does empirically in nearly all equilibrated liquids) then the terminal velocity of a dropped sphere in a liquid must be monotonically increasing in the energy density of the liquid. Thus, in the notation of Eq. (\ref{mc}), the (microcanonical ensemble averaged) equilibrium velocity $v'_{\infty}$ must be monotonic,  
 \begin{eqnarray}
\label{vvvv}
\frac{\partial v'_{\infty}(E')}{\partial E'}  = \frac{\partial}{\partial E'} \Big(\frac{ \sum_{E'- \Delta E \le E_{n} \le E'} \langle \phi_{n} | v_{z} | \phi_{n}\rangle}{{\cal{N}}[E'-\Delta E, E']}\Big) \ge 0.
\end{eqnarray}
Eq. (\ref{vvvv}) holds for any $E'$ and system size independent finite $\Delta E$. Thus, if localized states appear (for which the expectation value of the velocity vanishes) then they must be degenerate (or very nearly degenerate) in energy with states that are not localized. This degeneracy must be present to ensure that the average $v'_{\infty}(E')$ still grows as $E'$ is increased, notwithstanding the existence of such localized states in which no motion occurs. Richer possibilities may exist \cite{note_loc}.

\section{Low temperature glasses.}
\label{low-sec} 
From the constraint of Eq. (\ref{constraint}), at low temperatures only the ground and proximate low lying excited states have a measurable weight amplitude $|c_{n}|^{2}$ in Eq. (\ref{pt+}).
The system may then emulate two-level theories of low $T$ glasses \cite{2level,chandra,moshe} and related first principles approaches \cite{dervis}. Thus, our quantum theory, very naturally, links high and low temperature behavior in a general unified manner. In general, changes in the specific heat with $T'$ as well as crossovers in the probability density ${\it{p}}_{T}(E')$ with $T$  may lead to effective crossovers in the form for observables computed via Eqs. (\ref{thermog},\ref{stokes}, \ref{fix}) (and thus to departures from Eq. (\ref{geoni}) in the main text).

\section{Multi-scale probability densities.}
\label{multi-sec}
In the main text, we motivated the minimal Gaussian distribution consistent with Eq. (\ref{constraint}). This led to the single parameter (``$\overline{A}$'')
 fit of Fig. 1b  following 
 Eq. (\ref{geoni}). If $T$ is not the only temperature scale then, in principle, richer Gaussian (and other) probability distributions are possible. In this brief section, we would like to suggest that, when present, these additional temperatures may naturally lead to putative ``liquid-liquid'' transitions  \cite{ll}. 
 
Illuminating simulations \cite{itamar_also_change_at_TA} demonstrate that the effective free energy barriers may follow a bimodal distribution. Within our framework, these results would suggest that the probability density ${\it{p}}_{T}(E')$ may, similarly, be bimodal and contain two energy (or temperature) scales. We remark that one may justify such a distribution. For instance, if upon supercooling, (a) spatially jammed solid regions nucleate and grow and become progressively quiescent as $T$ is lowered while (b) the remaining unjammed fluid regions occupy a diminishing volume within which they retain their mobility then the probability distribution may be the sum of (a) a Gaussian at low energies or temperatures of weight $x(T)$ and (b) another Gaussian of weight $(1-x(T))$ with its support at energies near the melting energy. In the main text, we discussed the situation of $x=1$ and the only relevant temperature (scale) was $T$ itself, i.e., the average effective temperature in that case was $T_{1} =T$ (and the width of the distribution was set by $T$). In the above distribution, if at low temperatures, $x \to 1^-$ then, similar to the discussion in the main text, a dramatic rise of the viscosity with decreasing temperature (as in, e.g., Eq. (\ref{geoni})) may appear. Cross-overs in $x(T)$ may create the impression that phase transitions occur \cite{ll}. 
 By contrast, non-analyticities in $x(T)$ will lead to genuine transitions. Apart from the melting temperature scale $T_{melt}$, other natural lower temperature scales (e.g., the glass transition temperature $T_{g}$ and the (possibly related) temperature $T_{-}$ that was introduced in Section \ref{sec:Energy}) may appear if certain features ``freeze in'' at these temperatures and persist upon further supercooling.  
 From a different and more formal complimentary approach, whenever convex hull features appear \cite{zoltan}, bimodal and other distributions may appear instead of the standard single Gaussian of Eq. (\ref{Gauss'}) 
 or other forms found by direct maximization of the Shannon entropy.


\begin{thebibliography}{99}
\bibitem{define-glass} Commonly, the glass transition temperature $T_g$ is defined to be the temperature at which the viscosity is $10^{12}$ Pa $\cdot$ sec (or $10^{13}$ Poise).
\bibitem{ehrenfest} Rather explicitly, for a Hamiltonian which is the sum of a kinetic energy (of particles of mass $m$)
and a potential energy $V$, by the Ehrenfest theorem (or the expectation values of the equivalent Heisenberg or Schr\"odinger equations)
\begin{eqnarray}
\frac{d}{dt} \langle \psi| x_{\alpha} | \psi \rangle = \frac{\langle \psi | p_{\alpha} | \psi \rangle}{m}, \nonumber
\\ \frac{d}{dt} \langle \psi | p_{\alpha}| \psi \rangle = - \langle \psi| \frac{\partial V}{\partial x_{i\alpha}}| \psi \rangle. 
\label{Ehr.}
\end{eqnarray}
When the uncertainties in the particle positions $x_{\alpha}$ in the state $| \psi \rangle$ are small (as they typically are in semiclassical systems), then we may substitute 
\begin{eqnarray}
\langle \psi| \frac{\partial V}{\partial x_{\alpha}}| \psi \rangle =  \frac{\partial V}{\partial x_{\alpha}}|_{x_{\alpha} \to \langle \psi| x_{\alpha} | \psi \rangle}. 
\end{eqnarray}
With this, Eqs. (\ref{Ehr.}) reduce to the classical equations of motion.
In the quantum setting, the time evolution (from time $t$ to time $(t+ \Delta t)$) of the expectation values is given by $\langle \psi | x_{\alpha} | \psi \rangle  \to \langle \psi| e^{iH \Delta t/\hbar} x_{\alpha} e^{-iH \Delta t/\hbar} | \psi  \rangle,  ~ \langle \psi| p_{\alpha|} | \psi \rangle \to  \langle \psi| e^{iH \Delta t/\hbar}  p_{\alpha} e^{-iH \Delta t/\hbar} | \psi  \rangle$.

Similarly, if the dynamics in the classical system is subject to external random (thermal or other) forces $\eta_{\alpha}$ then Eq. (\ref{Ehr.}) is replaced by
\begin{eqnarray}
\label{aea}
\frac{d}{dt} \langle \psi| x_{\alpha} | \psi \rangle = \frac{\langle \psi | p_{\alpha} | \psi \rangle}{m} + \frac{\eta_{\alpha}}{m}, \nonumber
\\ \frac{d}{dt} \langle \psi | p_{\alpha}| \psi \rangle = - \langle \psi| \frac{\partial V}{\partial x_{i\alpha}}| \psi \rangle. 
\label{Ehr..}
\end{eqnarray}
If at time $t=0$ the classical and quantum expectation values are identical and if the noise $\eta_{\alpha}$ in the classical system coincides with the difference in the forces as measured in the quantum and classical systems,
i.e., if
\begin{eqnarray}
\label{eae}
\eta_{\alpha} = \Big( \frac{\partial V}{\partial x_{\alpha}} |_{x_{\alpha} \to \langle \psi| x_{\alpha} | \psi \rangle} \Big)- \langle \psi | \frac{\partial V}{\partial x_{\alpha}}| \psi \rangle,
\end{eqnarray} 
then the solutions $x_{\alpha}(t)$ and $p_{\alpha}(t)$ to the classical equations of motion with external noise (Eqs. (\ref{Ehr..})) will coincide with the measured observables of the corresponding position and momentum operators in the quantum state $| \psi(t) \rangle = e^{-iHt/\hbar} | \psi(0) \rangle$.
The noise $\eta$ in Eq. (\ref{Ehr..}) may indeed be thermal noise that assists a classical  supercooled liquid as it evolves according to the equations of motion to explore the ``energy landscape''. One may anticipate the magnitude of the ``uncertainties'' in position (triggered by external thermal noise) to become larger as the system is at higher temperature $T$. Since the Hamiltonian $H$ is a sum of position and momentum dependent terms, this anticipation 
of a larger uncertainty is in line with the larger effective broadening $\sigma_T$ that we will find for 
the energy eigenvalues in the state $| \psi \rangle$ at higher energy densities or, equivalently, higher
temperatures (see Eq. (\ref{wide'}) for a fixed $\overline{A}$ and effective average heat capacity $C$). For a classical system governed by Eq. (\ref{aea}) the energy uncertainty formally introduced by the random thermal agitation $\eta$ can be substantial; typically, this uncertainty has a scale set by the temperature.
\bibitem{e2+} For any function $F$ that (away from any phase space regions of localized support) is a piecewise linear (or constant) function in the each of the phase coordinate variables $z=x_i$ or $p_i$, 
we have that

\begin{eqnarray}
\langle  \psi | F(z) | \psi \rangle = F(\langle \psi | z| \psi \rangle).
\end{eqnarray}

Thus, for general functions $F = \partial H/\partial x_{i}$ or $F=\partial H/\partial p_{i}$ that are indeed linear in the phase space coordinates, Hamilton's classical equations of motion (which are those in  $F(\langle \psi | z| \psi \rangle)$ in which $\langle z \rangle$ (with $z=x$ or $z=p$)
are the classical phase space coordinates) are identical to those implied by Ehrenfest's theorem in the quantum arena.
In such circumstances, even if there are quantum uncertainties in any phase space variables, the classical equations of motion will still be obeyed
(following the replacement of the classical variables $x$ and $p$ by their quantum expectations values $\langle \psi |x| \psi \rangle$ and $\langle \psi| p |\psi \rangle$) whenever the quantum Hamiltonian $H$ is quadratic in the phase space variables. 
Whenever the wavefunction is localized about particular phase space coordinates, the corresponding position or momentum operators $z$ exhibit no fluctuations about their average values and the relation $\langle F(z)\rangle=F(\langle z\rangle)$ will, similarly, be ensured. That is, in such instances, Ehrenfest's theorem automatically implies the classical (i.e., Hamilton's) equations of motion are satisfied.
Solving any such system quantum mechanically or analyzing it classically will lead to identical results.
\bibitem{rejuv} Although we will not consider it in the current work, instead of Eq. (\ref{HHH}), one may also examine a single step process whereupon the initiation of cooling $H \to \tilde{H}$ (i.e., at all times $t>t_{initial}$ the Hamiltonian will be given by $\tilde{H} \neq H$). Similarly,
the evolution operator $ \tilde{U}$ may embody a ``rejuvenation'' process wherein the system is heated from the supercooled state.
\bibitem{note_loc} A priori, the expectation value of an observable such as the velocity operator (${\cal{O}} = v_{z}$) discussed in Section \ref{vlv},
$\langle \phi_{n} | v_{z} | \phi_{n}\rangle$ and other similar quantities, might not depend only on the energy $E_{n}$ of the eigenstate $|\phi_{n} \rangle$. Thus, as in Eq. (\ref{richer_dep}), the average in Eq. (\ref{vvvv}) may, theoretically, depend on quantum numbers ${\cal{Q}}$ other than the energy. Such a dependence may, in principle, distinguish between candidate localized and delocalized states that share the same energy density.
\bibitem{diagonalll} By the ``off-diagonal terms'' that have been omitted in the second equality of Eq. (\ref{fix}), we allude to the sum
\begin{eqnarray}
\label{upp}
\frac{d}{dt}  \sum_{n \neq n'} \sum_{m}  c_{n} c_{n'}^* \langle \phi_m|U_{pert.}(t)|\phi_{n} \rangle \langle \phi_m|U_{pert.}(t)| \phi_{n'} \rangle^* e^{-i(E_{n}-E_{n'})t/\hbar}.
\end{eqnarray}
Similar to the relations of Eqs. (\ref{llong}, \ref{thermog}) yet weaker, individual $n  \neq n'$ terms summed over in Eq. (\ref{upp}) may each have a vanishing long time average.
\bibitem{latent_vel} Strictly speaking, if the latent heat ${\cal{PT}}E{\cal{I}}$ region is included then Eq. (\ref{vv}) will hold for all eigenstates have an energy lower than the lowest energy of the ${\cal{PT}}E{\cal{I}}$. Within the ${\cal{PT}}E{\cal{I}}$ region, flow might occur. As such the relations that we will derive assuming that no flow occurs in the equilibrated system below the melting temperature $T_{melt}$ will lead to upper bounds on the viscosity (see, e.g., Eq. (\ref{boundratio})).
\bibitem{Gauss-remind} The derivation of this well-known relation is sketched here for the benefit of the reader. If we enforce a fixed variance $\sigma_T^2 = \int dE' (E'- E^2) \rho(E')$ with a Lagrange multiplier $\lambda_E$ while maximizing the Shannon entropy of Eq. (\ref{ME}) (and the condition of normalization $1=\int dE' \rho(E')$ with a Lagrange multiplier $\lambda_{1}$), the corresponding variational equation trivially reads $\ln \rho = (1- \lambda_{1}-\lambda_{T}(E'-E)^2)$. Demanding normalization and the fixed value of the standard deviation $\sigma_T^2$ fixes the Lagrange multipliers $\lambda_{T}$ and $\lambda_{1}$ and leads to Eq. (\ref{Gauss'}).
\bibitem{achtung} It is conceivable that distributions more complex than that of Eq. (\ref{Gauss''}) may appear when the supercooling is achieved in a very special way. Furthermore, the amended variant of Eq. (\ref{thermog}) (Eq. (\ref{richer_dep}) for general observables ${\cal{O}}$) may be of use in analyzing rare yet notable situations, see, e.g., \cite{ediger_new}, wherein different spatially inhomogeneous preparations of the supercooled state generally trigger significantly different behaviors.
\end{thebibliography}
\end{document}